\documentstyle[preprint,prd,aps]{revtex}

\begin{document}
\draft
\preprint{IFA-97/10}
\title{Partial-Wave Amplitudes and Resonances in $\overline{p}+p\rightarrow \pi
+\pi $ }
\author{B. R. Martin}
\address{Department of Physics and Astronomy, University College London, \\
London WC1E 6BT, England}
\author{G. C. Oades}
\address{Institute of Physics and Astronomy, Aarhus University,\\
DK-8000 Aarhus C, Denmark}
\date{submitted to Phys. Rev. D}
\maketitle

\begin{abstract}
Partial wave amplitudes have been extracted from accurate data on $\overline{%
p}p\rightarrow \pi ^{-}\pi ^{+}$, combined with earlier data on $\overline{p}%
p\rightarrow \pi ^{0}\pi ^{0}$, by a method which incorporates the
theoretical constraints of analyticity and crossing symmetry. The resulting
solution gives a good fit to the annihilation data and is also consistent
with the wealth of information in the crossed channel $\pi N\rightarrow \pi
N $. The partial wave amplitudes show evidence for resonances in all partial
waves with $J\leq 5$, at least one of which, a $J=0^{+}$ state, (and
possibly another with $J=1^{-}$) is unlikely to have a simple $q\overline{q}$
structure.
\end{abstract}

\pacs{13.75.Cs, 11.80.Et, 14.40.Cs, 25.43.+t}

\section{Introduction}

There have been a number of attempts to obtain information on meson
resonances from data on the reaction $\overline{N}N\rightarrow \pi \pi $.
Earlier work \cite{ref1} was restricted to testing the compatibility of data
with assumed combinations of Breit-Wigner resonances and smooth backgrounds.
These analyses used data only on the channel $\pi ^{-}\pi ^{+}$ \cite{ref2}
and the results, in the main, are in strong disagreement with later $\pi
^{0}\pi ^{0}$ data \cite{ref3}. An alternative approach is to use the
Barrelet zero method \cite{ref4} to translate the dips observed in the
angular distributions (cross-sections and asymmetries) into complex zeros of
the scattering amplitudes. This method requires a number of assumptions to
obtain smooth zero trajectories and to fix the overall phase of the
solution. The latter is crucial, because in the absence of an optical
theorem the method only determines the moduli and {\it relative} phases of
the amplitudes at each energy. One choice is to assume a Breit-Wigner form
for a particular partial wave, but different choices of input phases,
however plausible, lead to different solutions for the partial waves, as can
be seen by comparing the results of references \cite{ref5} and \cite{ref6}.

An alternative method of obtaining amplitudes which overcomes the objections
to earlier work, is to exploit analyticity via dispersion relations and
determine the overall phase by using crossing symmetry to relate the
annihilation data to known $\pi N\rightarrow \pi N$ elastic scattering
amplitudes. This in principle overcomes the problem of the absence of an
optical theorem and ensures that the resulting amplitudes not only fit the
annihilation data, but are also consistent with the wealth of information on
the crossed channel, $\pi N\rightarrow \pi N$. Dispersion relations at fixed-%
$t$ or fixed-$u$ are unsuitable because of the need to have information in
unphysical regions, but this can be avoided by writing dispersion relations
along hyperbolae in the Mandelstam plane \cite{ref7}. In a previous paper 
\cite{ref8} we presented a largely model-independent set of invariant
amplitudes, which by construction satisfy analyticity and crossing symmetry
and simultaneously give an excellent fit to all $\overline{p}p\rightarrow
\pi \pi $ data (both in charged and neutral channels) existing at the time 
\cite{ref2,ref3,ref9} and to amplitudes for $\pi N\rightarrow \pi N$.
Subsequently \cite{ref10} we showed that the structure in the invariant
amplitudes, which reflects that seen directly in the experimental data, was
due to the existence of resonances in the partial wave amplitudes and we
presented evidence for resonance activity in four states with spins $1,2,3$
and $4$.

Since our earlier work, new differential cross-section and asymmetry data
have been obtained on the $\pi ^{-}\pi ^{+}$ channel at 20 center-of-mass
energies in the range 1.91 MeV to 2.27 MeV \cite{ref11} using the LEAR
facility at CERN. These data are consistent with the earlier data \cite{ref2}%
, but are more accurate and also extend asymmetry measurements to lower
momenta than previous experiments. However, the new experiments cover a
slightly smaller range of energy. Here we present an analysis of the new
data, plus the older $\pi ^{0}\pi ^{0}$ data \cite{ref3}, using as
constraints the invariant amplitudes obtained in \cite{ref8}, thus
indirectly imposing analyticity and crossing symmetry on the solution. We
compare our results with other recent analyses \cite{ref12,ref13,ref14} at
the end of this paper.

\section{Partial-Wave Amplitudes}

The analysis of \cite{ref8} was performed in terms of the usual $\pi N$
invariant amplitudes $A$ and $B$. In the $t$-channel, $\overline{N}%
N\rightarrow \pi \pi $, it is more convenient to work with $B$ and 
\begin{equation}
C\equiv -A+M\left( \frac{q}{p}\right) \cos \theta _{t}B,  \label{eq1}
\end{equation}
where $q(p)$ is the center-of-mass momentum in the $\pi \pi (\overline{N}N)$
channel; $M$ is the nucleon mass; and $\theta _{t}$ is the scattering angle
in the $t$-channel. In the helicity basis, these two invariant amplitudes
may be expressed in terms of partial-wave amplitudes by the expansions 
\begin{equation}
B^{I}(t,\cos \theta _{t})=\frac{8\pi }{p}\sum_{J}\frac{2J+1}{\sqrt{J(J+1)}}%
P_{J}^{\prime }\left( \cos \theta _{t}\right) F_{J}^{I}\left( t\right)
\label{eq2}
\end{equation}
and 
\begin{equation}
C^{I}\left( t,\cos \theta _{t}\right) =\frac{4\pi \sqrt{t}}{p}\sum_{J}\left(
2J+1\right) P_{J}\left( \cos \theta _{t}\right) N_{J}^{I}\left( t\right) .
\label{eq3}
\end{equation}
In these relations, $N_{J}^{I}$ and $F_{J}^{I}$ are the partial-wave
helicity amplitudes for definite $t$-channel isospin $I$ corresponding to $t$%
-channel helicity non-flip and flip, respectively, and $t$ is the square of
the center-of-mass energy in the $t$-channel. By Bose statistics, 
\begin{equation}
N_{J}^{1}(t)\equiv F_{J}^{1}(t)\equiv 0\qquad \text{for }J\text{ even}
\label{eq4}
\end{equation}
and 
\begin{equation}
N_{J}^{0}(t)\equiv F_{J}^{0}(t)\equiv 0\qquad \text{for }J\text{ odd.}
\label{eq5}
\end{equation}
In addition, $F_{0}^{0}\equiv 0$. The helicity amplitudes are normalized
such that the integrated cross-section for $\overline{N}N\rightarrow \pi \pi 
$ for a definite isospin $I$ is 
\begin{equation}
\sigma ^{I}=2\pi \left( \frac{q}{p}\right) \sum_{J}\left( 2J+1\right)
\left\{ \left| N_{J}^{I}\right| ^{2}+\left| F_{J}^{I}\right| ^{2}\right\} .
\label{eq6}
\end{equation}
An alternative, but equivalent, way of expressing the partial-wave content
of the invariant amplitudes is to use the $LS$ basis. If we label the new
amplitudes by $JL$, i.e. $H_{JL}$, then, for a fixed value of isospin $I$,
they are related to $N_{J}$ and $F_{J}$ by 
\begin{equation}
H_{J,L=J-1}=\left( \frac{2}{2J+1}\right) ^{1/2}\left\{ \sqrt{J+1}F_{J}-\sqrt{%
J}N_{J}\right\}  \label{eq7}
\end{equation}
and 
\begin{equation}
H_{J,L=J+1}=\left( \frac{2}{2J+1}\right) ^{1/2}\left\{ \sqrt{J}F_{J}+\sqrt{%
J+1}N_{J}\right\}  \label{eq8}
\end{equation}
with 
\begin{equation}
\sigma =\pi \left( \frac{q}{p}\right) \sum_{J}\left( 2J+1\right) \left\{
\left| H_{J+}\right| ^{2}+\left| H_{J-}\right| ^{2}\right\} ,  \label{eq9}
\end{equation}
where $H_{J\pm }\equiv H_{J,L=J\pm 1}$. We will use both sets of amplitudes
in the following discussion.

In\cite{ref8}, the invariant amplitudes were obtained along families of
hyperbolic curves in the Mandelstam plane defined by a parameter 
\begin{equation}
\xi \equiv \frac{\left( M^{2}-m^{2}\right) ^{2}-s\left( \Sigma -s-t\right) }{%
t-4m^{2}}  \label{eq10}
\end{equation}
where $m$ is the pion mass; $s$ is the square of the center-of-mass energy
in the $s$-channel $\pi N\rightarrow \pi N$; and $\Sigma \equiv 2\left(
M^{2}+m^{2}\right) $. Values of $\xi $ were chosen so that the hyperbolae
covered the $t$-channel region being analysed while staying almost entirely
within the physical $s$-channel. To obtain amplitudes at the experimental
energies, those from ref. \cite{ref8} were linearly interpolated in $t$
along the hyperbolae at fixed-$\xi .$

\section{Fits to data and amplitudes}

The data fitted consisted of experimental differential cross-sections and
asymmetries for $\overline{p}p\rightarrow \pi ^{-}\pi ^{+}$ at 20 momenta in
the range 360 MeV/c to 1550 MeV/c \cite{ref11} and, at each momentum, the
invariant amplitudes from ref.\cite{ref8} at a grid of $\xi $-values (or
equivalently, a grid of values of $\cos \theta _{t}$). In addition,
differential cross-sections for $\overline{p}p\rightarrow \pi ^{0}\pi ^{0}$
were included by interpolating the measured data of ref.\cite{ref3} to the
momenta of the charged channel data. At each momentum we minimized the sum
of the values for $\chi ^{2}$ per data point for each type of data. The
parameters are, at each $t$-value, the complex amplitudes $N_{J}^{I}$ and $%
F_{J}^{I}$. The number of amplitudes used in the partial-wave expansions is
dictated by the form of the invariant amplitudes and the need to ensure a
good representation of them, and not by criteria such as polynomial fits to
the measured angular distributions, which are often truncated where the
highest partial wave is still quite large. This is an important difference
between our method and others \cite{ref5,ref6,ref12,ref13} which use
observables in the form of Legendre series truncated at a point where the
size of the highest coefficient is actually large. Our interpolation of the
t-channel data used a method that makes no assumption about the number of
partial waves which are important (see Appendix C of ref. \cite{ref8}),
although we would expect higher partial waves to be progressively less
important at lower momenta.

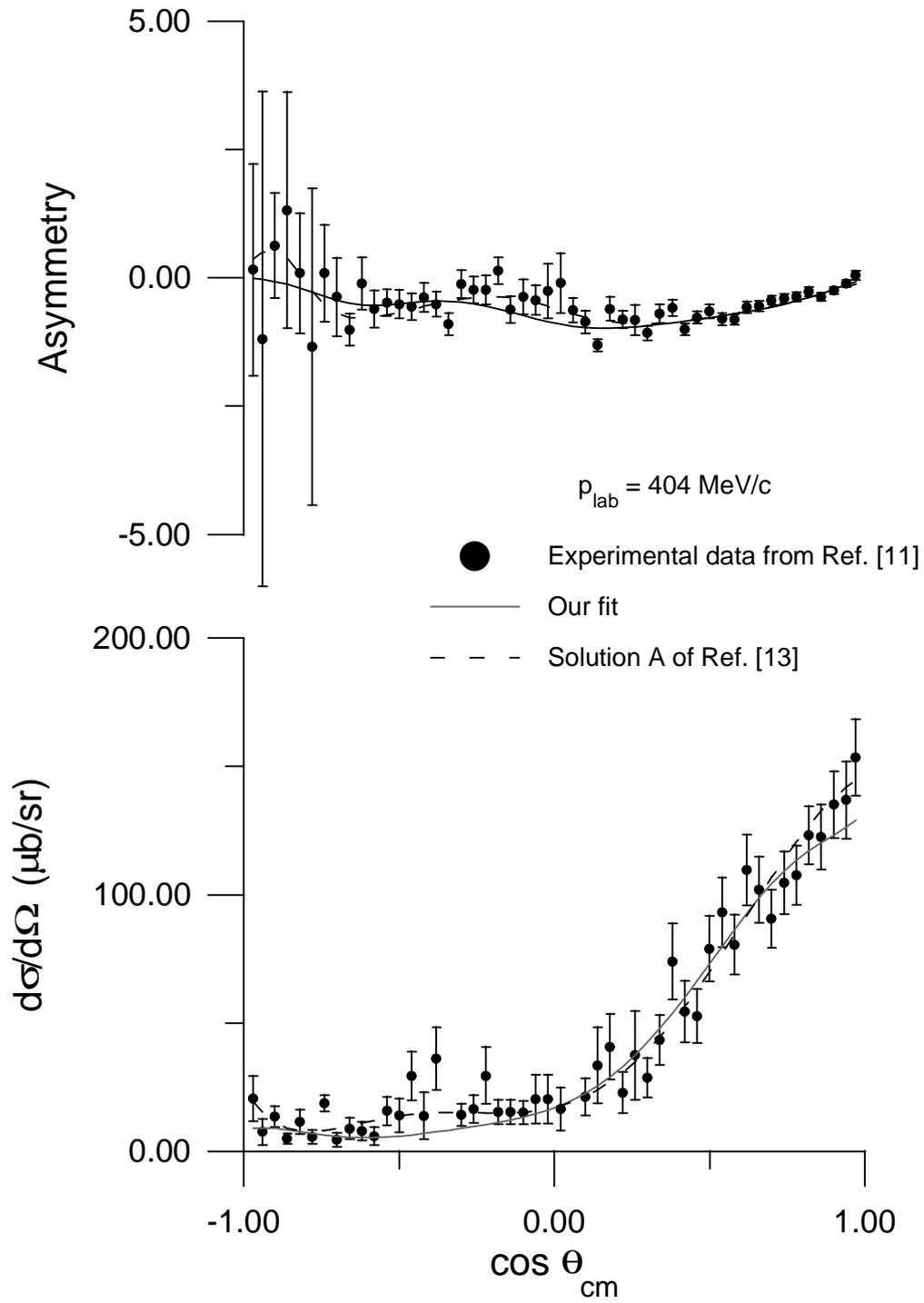
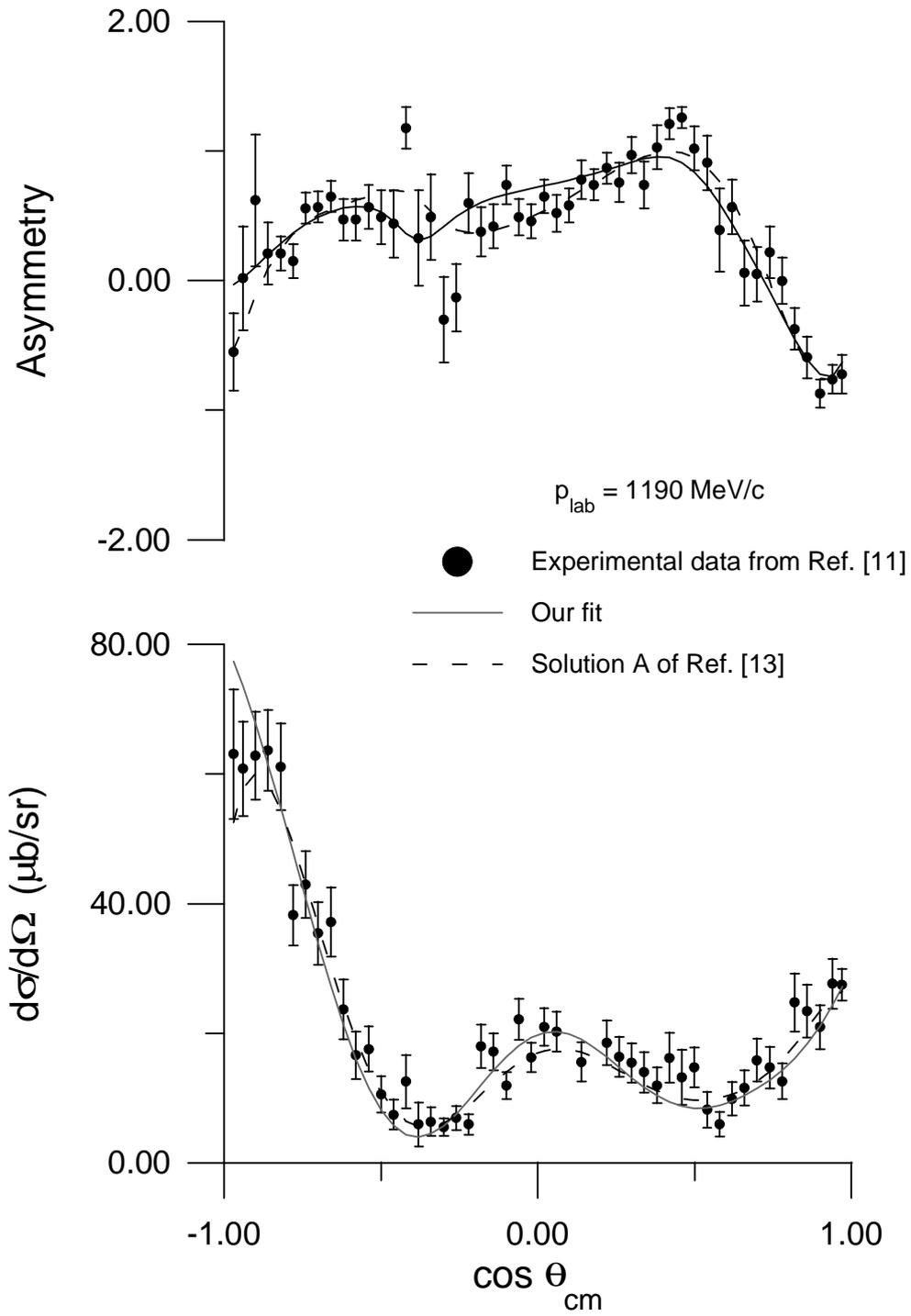
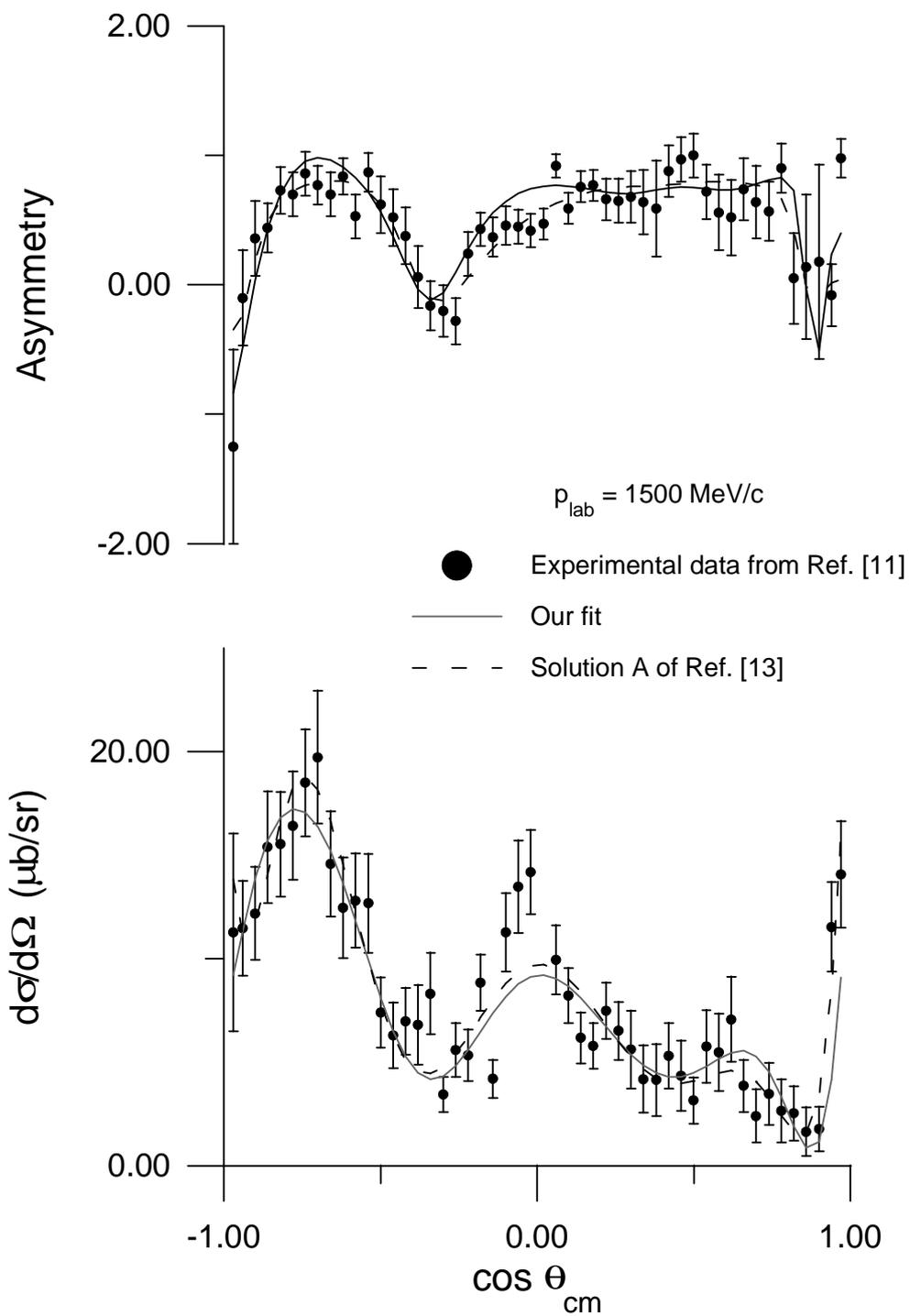
\begin{figure}[tbp]
\caption{Fits to the measured differential cross-section and asymmetry data
for $\bar{p}p \rightarrow \pi ^{-} \pi ^{+}$. Our results are shown by solid
lines and those of solution A of ref. \protect\cite{ref13} by dashed lines
(solution B is very similar).(a) 404 MeV/c, (b) 1190 MeV/c, (c) 1500 MeV/c.}
\label{fig1}
\end{figure}

At each $t$-value, sufficient terms were used to obtain a good fit to each
amplitude and to ensure reasonable smoothness of the resulting helicity
amplitudes from one $t$-value to the next. In practice, 4-5 $J$-values were
used at the lowest momentum, rising to 6-7 at the highest. Initial estimates
of $N_{J}^{I}$ and $F_{J}^{I}$ were obtained from \cite{ref8} and used as
starting values in a simultaneous fit to all the data at a given momentum.
In Table \ref{table1} we show values of $\chi ^{2}$ per data point for the
differential cross-sections and asymetries in the charged channel and for
the amplitudes at each of the twenty momenta. At the higher momenta we also
show values of $\chi ^{2}$ per data point for the neutral data. In Figs \ref
{fig1} and \ref{fig2} we show the fits and/or predictions for the $\pi
^{+}\pi ^{-}$ and $\pi ^{0}\pi ^{0}$ data at three representative momenta.
To give some idea of the quality of the fits to the invariant amplitudes, we
show in Fig. \ref{fig3} the fits to the real and imaginary parts of $B^{\pm
} $ and $C^{\pm }$ as functions of $t$ at a fixed value of the hyperbolic
parameter, $\xi =-0.04572$ GeV$^{2}$.

\section{Resonances}

The extraction of ``hard'' information on resonances from partial-wave
amplitudes is a notoriously difficult problem, which has not been solved in
a rigorous way even for elastic two-body reactions, and ultimately one has
to resort to plausible models. We will use criteria which have proved
successful in analysing reactions such as $\pi N\rightarrow \pi N$. Thus, we
will use combinations of Breit-Wigner resonances and flexible non-resonant
backgrounds with initial parameters suggested by loops in Argand diagrams,
enhancements in the integrated partial cross-sections, maxima in amplitude
speed plots etc., not by fitting the original experimental data. Before
doing that, however, it is useful to see to what extent the ``continuity''
of the partial-wave amplitudes, which were obtained by a series of fits at
fixed values of the energy, is compatible with partial-wave analyticity. To
test this we have fitted the ``raw'' amplitudes with the following
parametric forms constructed by analogy with the work of ref.\cite{ref8}.

\begin{equation}
H_{J,L=J\pm 1}=(1-z_{+})^{L}\;(1+z_{-}{\bf )}^{2}\left[
\sum_{n=0}^{N_{-}}a_{n}\;z_{-}^{n}\text{ }+\sum_{n=1}^{N_{+}}b_{n}%
\;z_{+}^{n}+\sum_{n=1}^{N_{u}}c_{n}\;z_{u}^{n}\right] ,  \label{eq10a}
\end{equation}
where the variables $z_{-}$ , $z_{+}$ and $z_{u}$ are given by 
\begin{equation}
z_{-}=\frac{a_{-}-(t-t_{L})^{1/2}}{a_{-}+(t-t_{L})^{1/2}},  \label{eq10b}
\end{equation}
\begin{equation}
z_{+}=\frac{a_{+}-(4\;M^{2}-t)^{1/2}}{a_{+}+(4\;M^{2}-t)^{1/2}}
\label{eq10c}
\end{equation}
and 
\begin{equation}
z_{u}=\frac{a_{u}-(4\;m^{2}-t)^{1/2}}{a_{u}+(4\;m^{2}-t)^{1/2}}.
\label{eq10d}
\end{equation}
Here we use the values $a_{-}$ $=$ $a_{u}$ $=$ 1 GeV and $a_{+}$ $=$ 2 GeV
for the mapping constants. In eq.\ref{eq10b} $t_{L}$ is given by 
\begin{equation}
t_{L}=m^{2}(4-\frac{m^{2}}{M^{2}}).  \label{eq10e}
\end{equation}
The three terms in eq.\ref{eq10a} ensure that $H_{J,L=J\pm 1}$ has the
correct analyticity properties as a function of $t$. The factor $%
(1-z_{+})^{L}$ ensures the correct behaviour at the $\overline{N}N$
threshold while the factor $(1+z_{-})^{2}$ ensures a suitable high energy
behaviour. We do not attempt to impose the correct behaviour at the $\pi \pi 
$ pseudo-threshold since this is very far away from the energy region in
which we are interested. The advantage in using such an a representation in
the $JL$ basis is that the $\overline{N}N$ threshold behaviour can be easily
imposed, whereas in the helicity basis this is more difficult since the
threshold behaviours of the two helicity states are correlated. The price
for this decoupling is a correlation between the two $JL$ basis amplitudes
at $t=0$ to avoid a spurious singularity at this point. However, this is
again so far from the energy region in which we are interested that we
ignore this problem.

The coefficients $a_{n}$ , $b_{n}$ and $c_{n}$ in eq.\ref{eq10a} are
determined by fitting the values of $H_{J,L=J\pm 1}$ at the different $t$%
-values using the Pietarinen technique as described in ref. \cite{ref8}. In
these fits we use $N_{-}=N_{u}=10$ and $N_{+}=15$ and are able to achieve
good fits; it should be remarked that in these fits we take no account of
correlations, neither between different amplitudes nor between the real and
imaginary parts in the same amplitude. The results of these fits are shown
as the solid curves in Fig. \ref{fig4} for $J\leq 5$ where we show the
dimensionless amplitudes 
\begin{equation}
h_{J\pm }(W)=(pq)^{1/2}H_{J,L=J\pm 1}(W).  \label{eq11}
\end{equation}
The $J=6$ amplitudes are small and featureless and so are not shown. We
conclude that, in the main, the single-energy amplitudes are compatible with
partial-wave analyticity within plausible errors. In what follows, we have
used these smooth amplitudes as the starting point for our extraction of
resonance information. This is done for convenience, and it is worth
remarking that no extra structure is introduced by the smoothing procedure
and that we have checked that our conclusions are unaltered when the ``raw''
amplitudes are fitted.

In Fig. \ref{fig4} we also show the contributions of each partial-wave
helicity amplitude to the integrated partial cross-section. Some systematic
features are immediately apparent. Firstly, in a given energy region odd-$J$
contributions are larger than those with even-$J$. This is of course
directly related to the experimental observation that the $\pi ^{0}\pi ^{0}$
cross-section is approximately 1/3 of the $\pi ^{-}\pi ^{+}$ cross-section
over the whole of the energy range we consider. We also note the increasing
importance of partial waves with lower $L$ as the energy decreases. Many
partial waves show counterclockwise loops, with the amplitude moving rapidly
over some part of the energy range; the classic signal of resonance
activity. To obtain resonance parameters, we have fitted the dimensionless
amplitudes with the parametric forms 
\begin{equation}
h_{J\pm }(W)=\frac{\alpha _{J\pm }}{M_{R}-W-i\Gamma /2}+\widetilde{p}%
^{L+1/2}\sum_{n=1}^{n_{JL}}\beta _{J\pm }^{(n)}x^{n-1}  \label{eq12}
\end{equation}
where $W\equiv \sqrt{t}$ and 
\begin{equation}
x\equiv \frac{2W-W_{\min }-W_{\max }}{W_{\max }-W_{\min }}.  \label{eq13}
\end{equation}
In the background term the coefficients $\beta _{J\pm }^{(n)}$ are complex
parameters, and to ensure the correct behaviour at the $\overline{N}N$
threshold we set $\widetilde{p}=p/p_{B}$ where $p_{B}$ is a suitable
momentum, in practice taken to correspond to $W=2.1$GeV. In the resonance
term, the parameters are the mass $M_{R}$, the width $\Gamma $ and the
complex residue $\alpha _{J\pm }$. To ensure the correct behaviour at the $%
\overline{N}N$ threshold, we set 
\begin{eqnarray}
\alpha _{J\pm } &=&\gamma _{J\pm }(\frac{p}{p_{R}})^{L+1/2},\;\;\;p\leq p_{R}
\label{eq14} \\
&=&\gamma _{J\pm },\;\;\;\;\;\;\;\;\;\;\;\;\;\;\;p>p_{R}  \nonumber
\end{eqnarray}
where $p_{R}$ is the value of $p$ at $W=M_{R}$ and $\gamma _{J\pm }$ is a
complex constant. Both amplitudes for a given $J$ were fitted
simultaneously, and in the absence of a realistic error analysis equal
weights were assigned to each point. The results obtained by fitting the
amplitudes as functions of energy, and the associated Argand diagrams, are
shown as dashed lines in Fig. \ref{fig4} where it will be seen that the fits
are as good as those using the more general parametrization of eq.\ref{eq10a}

The $J=0$ amplitude shows a clear resonance loop corresponding to a mass of
1.95 GeV/c$^{2}$ and a width 0.16 GeV. The resonance dominates the real part
at low energies, but there is a large background at the upper end of the
range. The imaginary part has significant contributions from both resonance
and background terms throughout the entire energy range. In the case of the $%
J=1$ amplitudes, there is a clear resonance signal in the $L=0$ amplitude
corresponding to a mass of 1.97 GeV/c$^{2}$ and a width of 0.14 GeV. This is
accompanied by a substantial background contribution. The $L=2$ amplitude is
by comparison much smaller. For $J=2$, both $h_{2-}$ and $h_{2+}$ amplitudes
have resonance behaviour. The mass is 1.93 GeV/c$^{2}$ and the width is 0.15
GeV. The $J=3$ state has a resonance in the $h_{3-}$ amplitude with a mass
of 2.02 GeV/c$^{2}$ and a width of 0.22 GeV. The width in the case of the
raw amplitudes is slightly larger at 0.26 GeV. The $h_{3+}$ amplitude is
very small. The resonance in the $J=4$ wave, which again is more prominent
in the $L=J-1$ amplitude, has a mass of 2.02 GeV/c$^{2}$ and a width of 0.14
GeV, which rises to 0.26 GeV in the case of the raw amplitudes. The mass is
in good agreement with that of the well-established f$_{4}$(2050) resonance,
and the values for the width span the accepted value of 0.21 GeV. The fact
that this state emerges clearly from our analysis lends weight to the
validity of our procedures and the parameters of the other predicted
resonances. In the $J=5$ wave there is evidence for a state at 2.19 GeV/c$%
^{2}$ with a width of 0.22 GeV, again coupling stronger to the $L=J-1$
amplitude. Finally, there is some evidence for a broad resonance in $J=6$,
but as this is just outside the range of the analysis, and the raw
amplitudes shown much more scatter, we do not comment on this possibility
further and it is therefore not shown.

The resonance masses and widths are summarised in Table \ref{table2}. Where
two values are given, the first corresponds to fiting the raw amplitudes and
the second to fiting the smoothed amplitudes. In the other cases both values
are the same. Table \ref{table2} also gives the values of the product of the
branching ratios 
\begin{equation}
B_{J}\equiv B(R\rightarrow \pi \pi )B(R\rightarrow \overline{N}N),
\label{eq15}
\end{equation}
calculated from the values of the residue parameters $\gamma _{J\pm }$. The
only case where there is a well-established resonance is the $J^{P}=4^{+}$
state f$_{4}$(2050) with a measured mass of 2.044$\pm $0.011 GeV/c$^{2}$ and
a width of 0.208$\pm $0.013 GeV. If we identify our $J=4$ state with this,
we can use the known $\pi \pi $ branching ratio of the f$_{4}$(2050) of 17\%
to estimate the $\overline{N}N$ branching ratio to be between 2\% and 8\%,
which is not inconsistent with a conventional $\overline{q}q$ resonance. If
we assume that $\pi \pi $ branching ratios in the range 10-20\% are also
typical of daughter states in this mass range, then we can see from Table 
\ref{table2} that the other predicted resonances with $J=2,3$ and $5$ have $%
\overline{N}N$ branching ratios similar to that of the f$_{4}$(2050).
However, the states with $J=0$ and $1$ are different and have far larger
branching ratios. Indeed for the $J=0$ state the coupling is probably
unreasonably large, but in this case there is a very large background
accompanying the resonance and in this situation our simple parameterisation
may not be appropriate to extract accurate resonance parameters. Such states
are unlikely to be conventional $\overline{q}q$ mesons, but are more likely
to have a multiquark $\overline{qq}qq$ structure \cite{ref15}.

\section{Other analyses and conclusions}

Three other analyses \cite{ref12,ref13,ref14} of the accurate $\pi ^{-}\pi
^{+}$ data \cite{ref11} have been published.In ref. \cite{ref12} it is
assumed from the outset that the data can be fitted by partial-wave
amplitudes each of which is parameterised in terms of towers of nearly
degenerate resonances. The analysis then fits all the data \cite
{ref2,ref3,ref11} simultaneously. The motivation for this assumption is
based on qualitative features of the data, such as the persistence of a very
large polarization over a wide energy range, but nevertheless is really only
testing the compatibility of the data with an assumed parametric form, like
analyses of earlier data \cite{ref1}. The resulting speed plots show little
evidence for resonance-like behaviour and only the $J=4$ and $5$ amplitudes
show clear peaks. This illustrates the strong correlations that exist
between the states in the various towers and makes it difficult to ascribe
much reliability to the parameters of individual resonances. This is further
illustrated by the predicted values of $B_{J}$ which, with the sole
exception of the $J=4$ state, where the mass and width were fixed from
experiment, all indicate very large branching ratios to the $\overline{N}N$
channel.

The analysis of ref. \cite{ref13} is based on the Barrelet method \cite{ref4}
and therefore has the usual problems intrinsic to that method associated
with the unresolved phase. The initially very large number of possible
solutions is reduced by using the $\pi ^{0}\pi ^{0}$ data \cite{ref3} and
further reduced by imposing threshold behaviour on the way zeros appear,
leaving just two solutions (A and B), both of which exhibit resonances which
for (B) are in all waves from $J=0$ to $J=4$.

Finally, a simple amplitude analysis has been made \cite{ref14} which finds
no compelling evidence for any resonances. However, the analysis only used
data in the restricted momentum range of 360 MeV/c to 988 MeV/c and is
therefore not strictly comparable with the other analyses reported here.

The work of ref. \cite{ref13} is the most complete of the published
analyses, but it is difficult to make a meaningful comparison with our own
because of the significant difference in the input data to these two
analyses i.e. our inclusion of invariant amplitudes and, via these, the
requirement that the solution should be consistent with information in the $%
\pi N\rightarrow \pi N$ channel. To see the importance of this, we show in
Fig. \ref{fig3} the prediction of the solutions of ref. \cite{ref13} for the
invariant amplitudes as functions of $t$ at the fixed value of $\xi
=-0.04572 $ GeV$^{2}$ together with our own fit to these amplitudes. It is
clear, that despite the considerable errors on the invariant amplitudes,
they do provide important constraints which are not satisfied by either of
the solutions of ref. \cite{ref13}.

Nevertheless, there are some common results found in both analyses. Both
find a clear signal for the f$_{4}$(2050), and both find its coupling to the 
$L=3$ amplitude is much stronger than to $L=5$. Also, both analyses yield a
low-lying $J=0$ state with a suggestion of abnormally large coupling to the $%
\overline{N}N$ channel. Agreement for other waves is much less good. For
example, for $J=1$ ref. \cite{ref13} finds a resonance only in solution B
and it couples mainly to $h_{1+}$, whereas the $J=1$ resonance in our
solution, although it has similar parameters, couples mainly to $h_{1-}$.
Likewise, there are substantial differences in masses, widths and couplings
for the $J=2$ and $3$ waves.

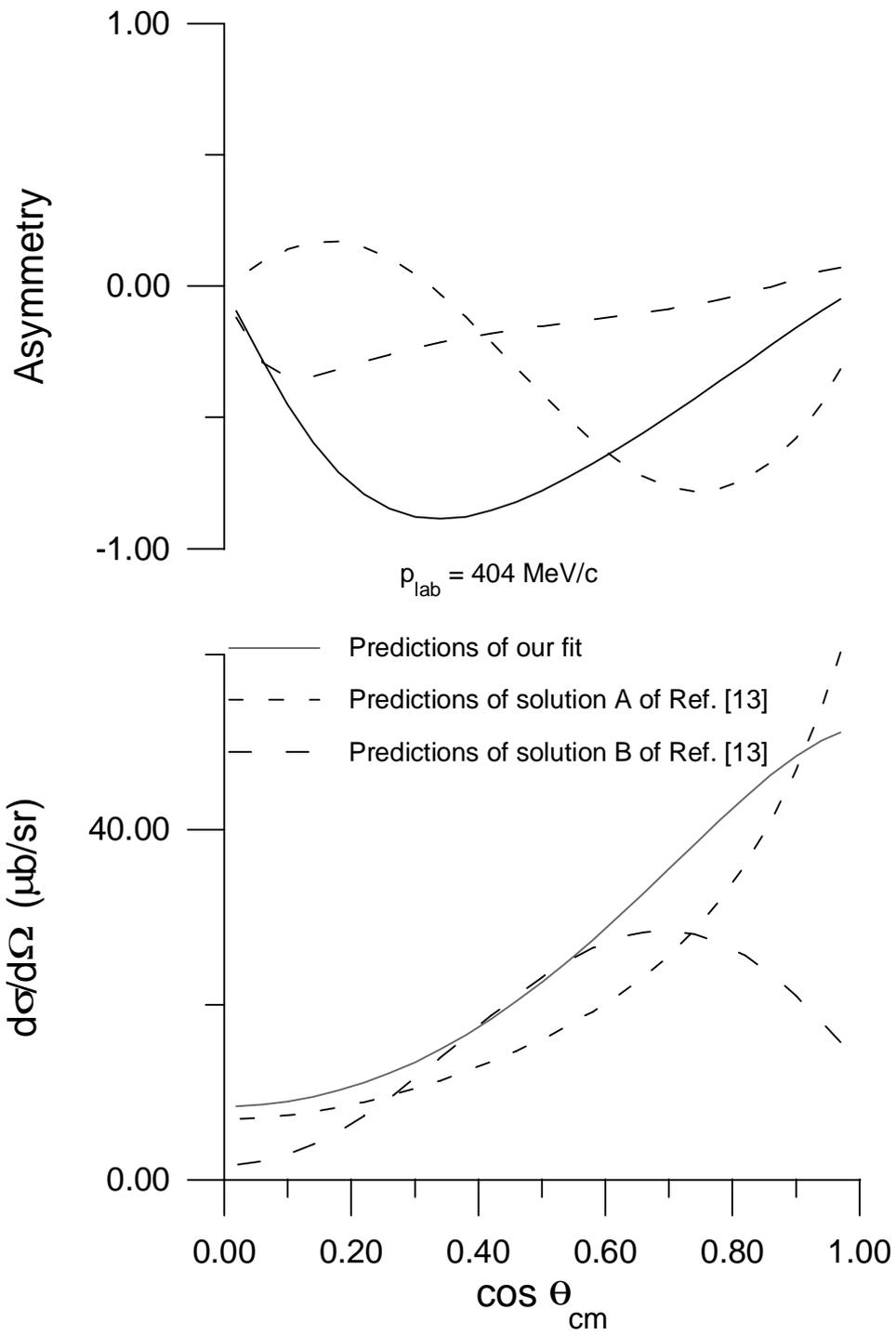
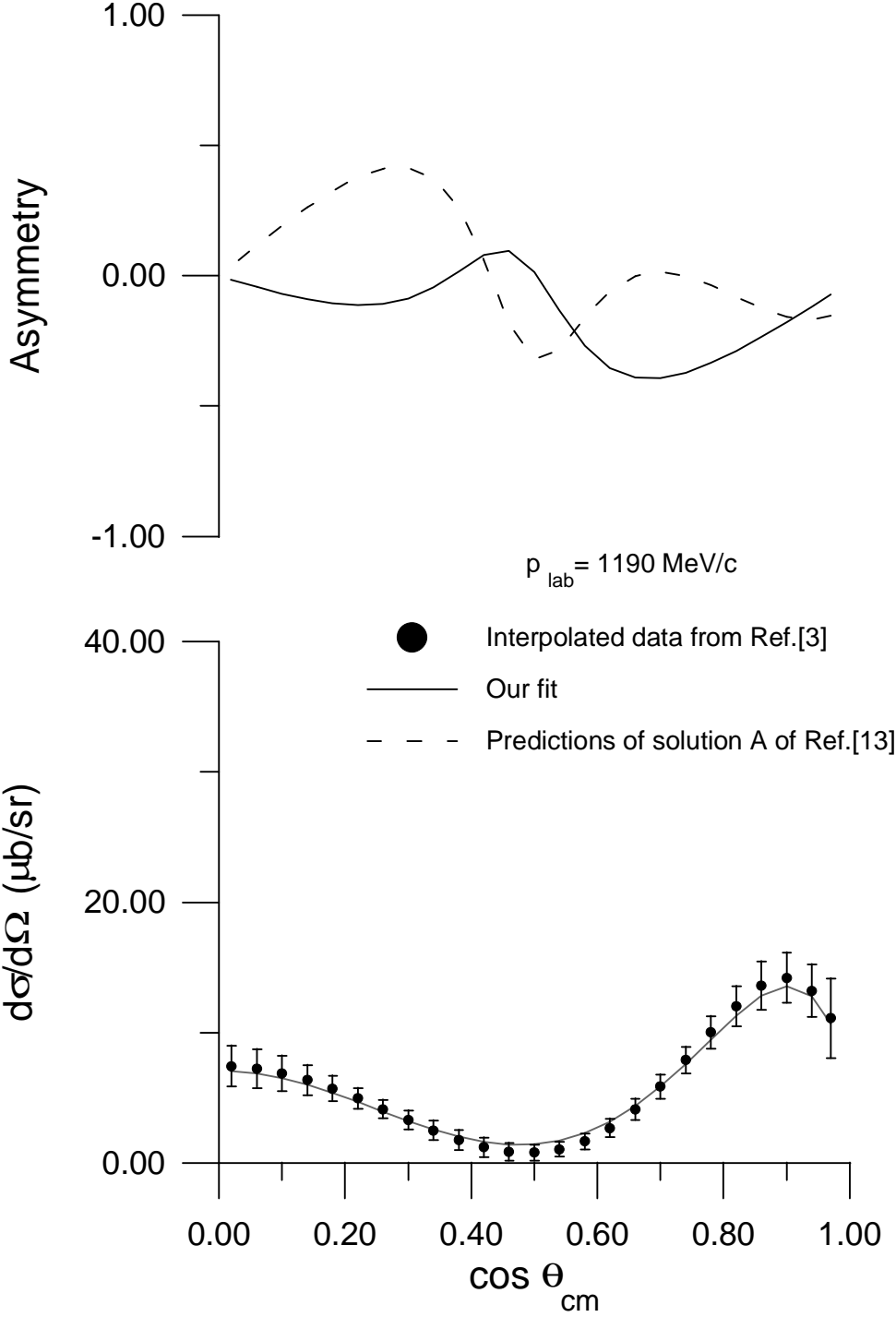
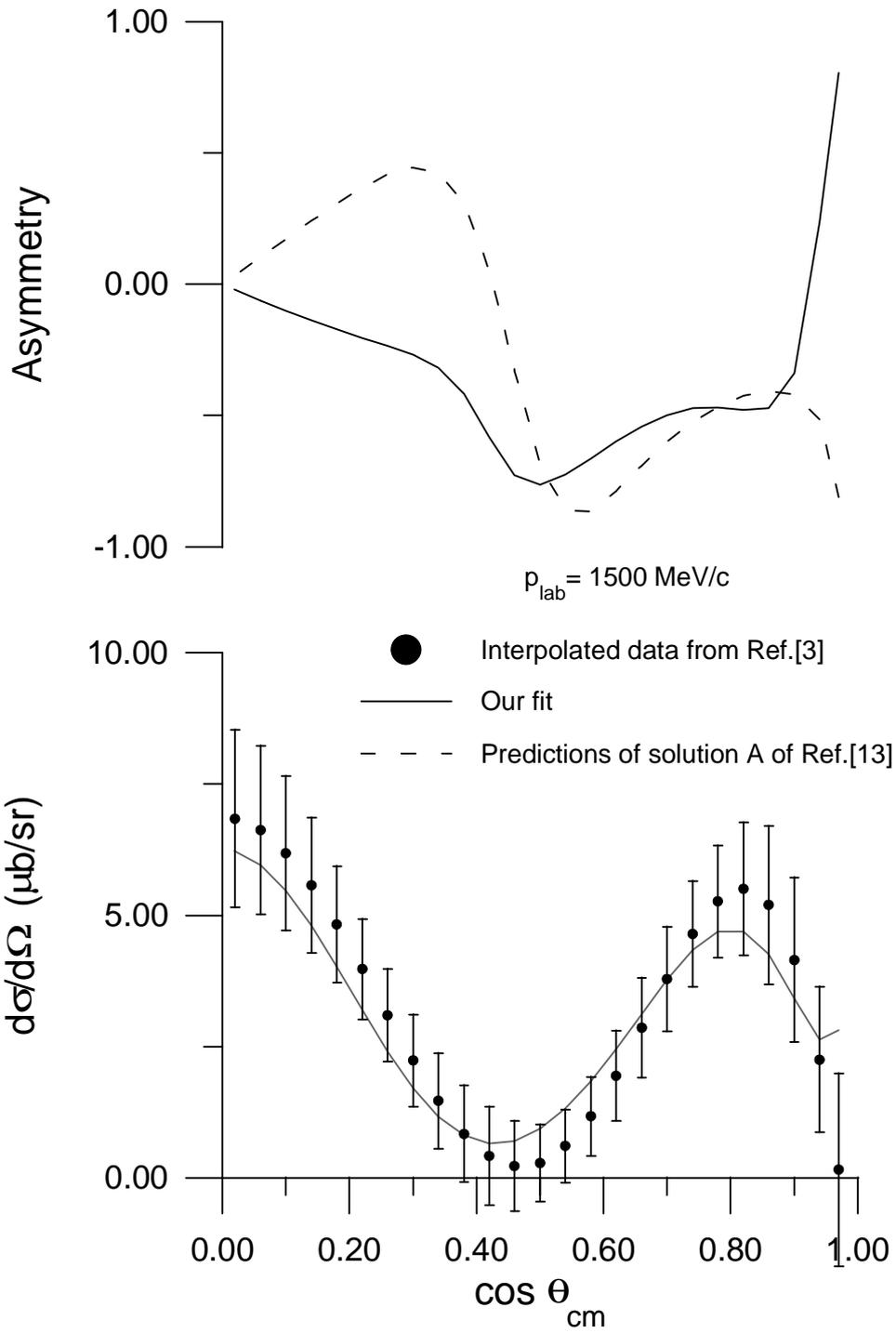
\begin{figure}[tbp]
\caption{Predictions and/or fits to the measured differential cross-section
and asymmetry data for $\bar{p}p \rightarrow \pi ^{0} \pi ^{0}$. Our results
are shown by solid lines and those of ref. \protect\cite{ref13} by dashed
lines. (a) 404 MeV/c, (b) 1190 MeV/c, (c) 1500 MeV/c.}
\label{fig2}
\end{figure}

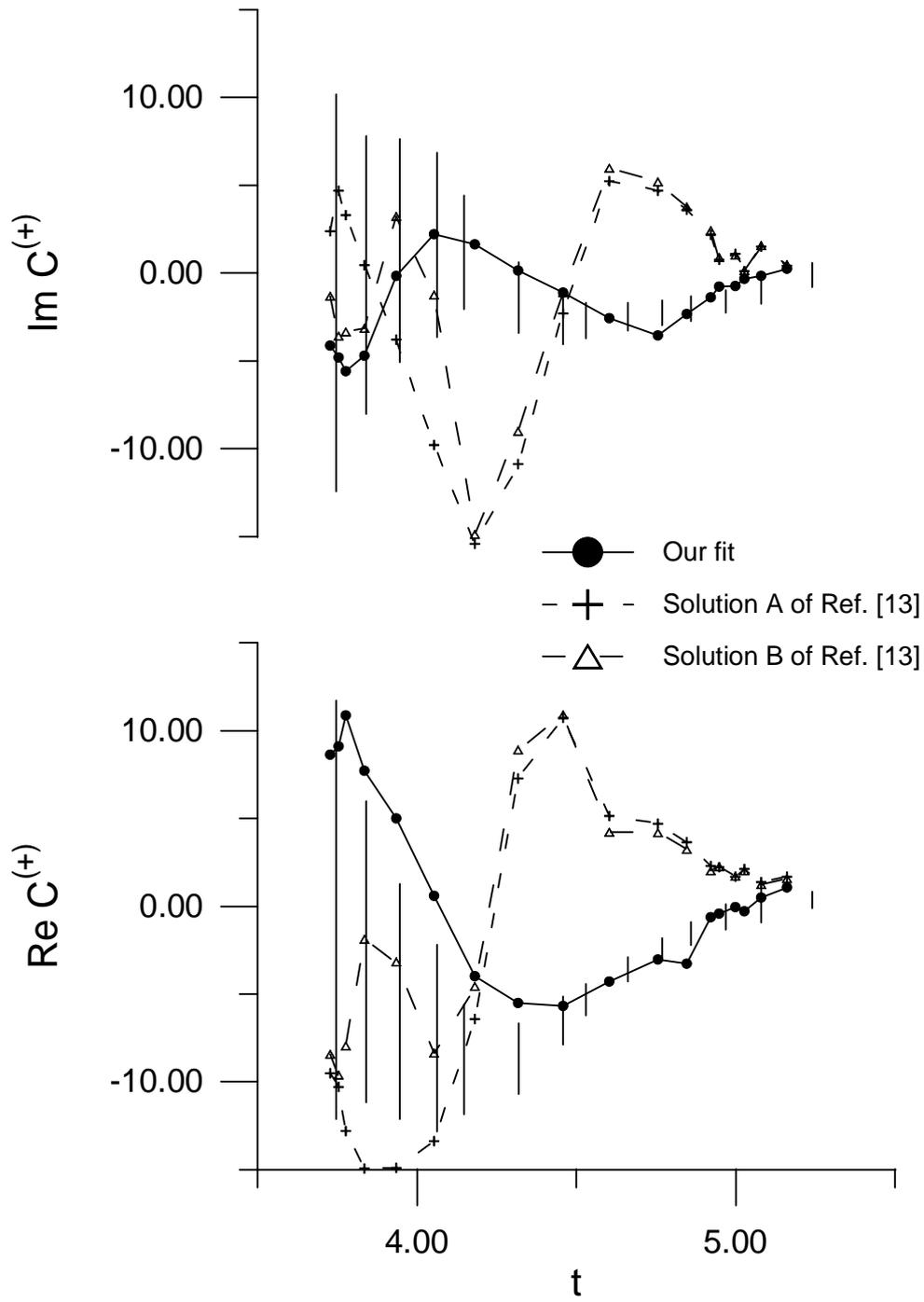
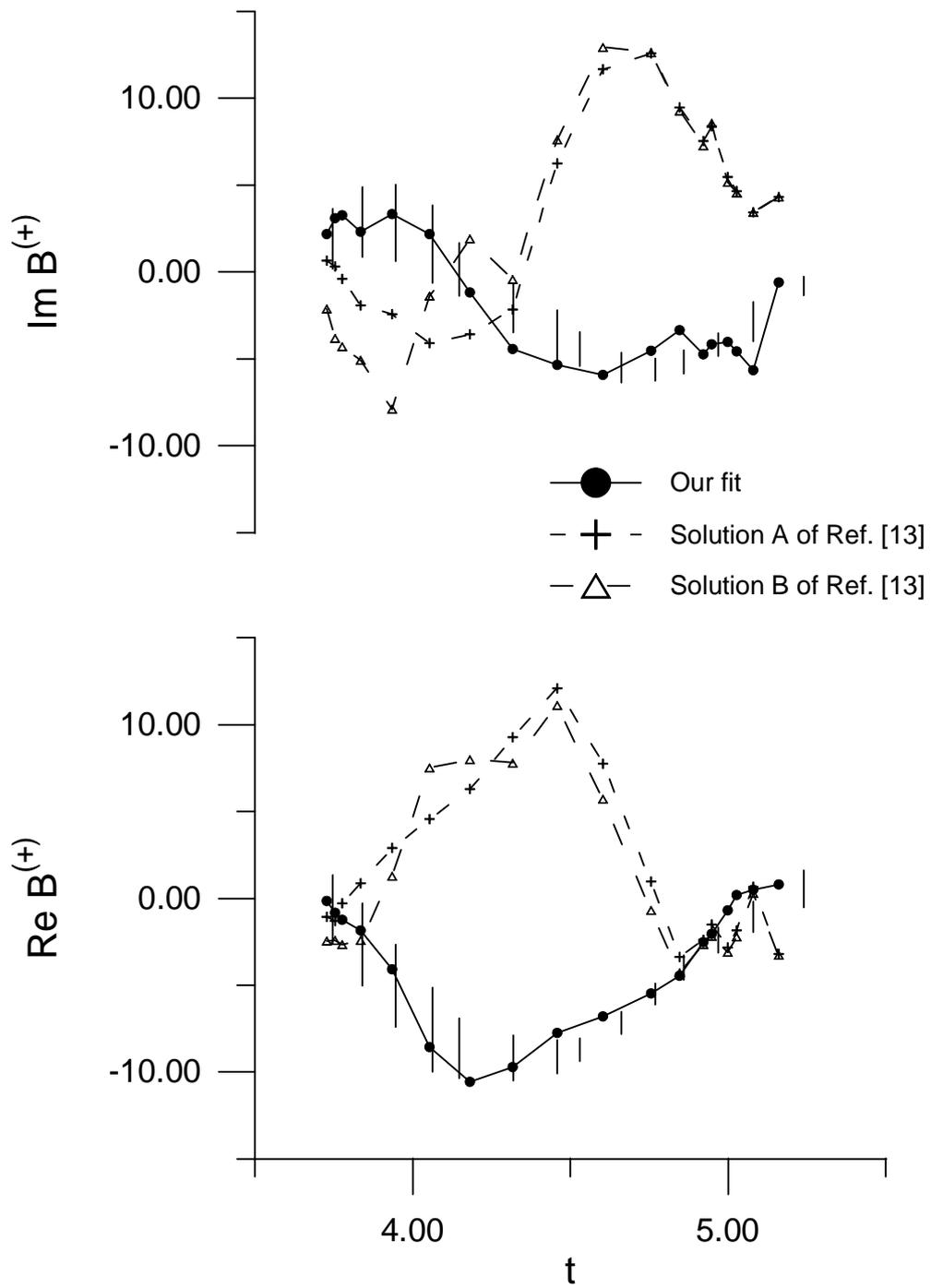
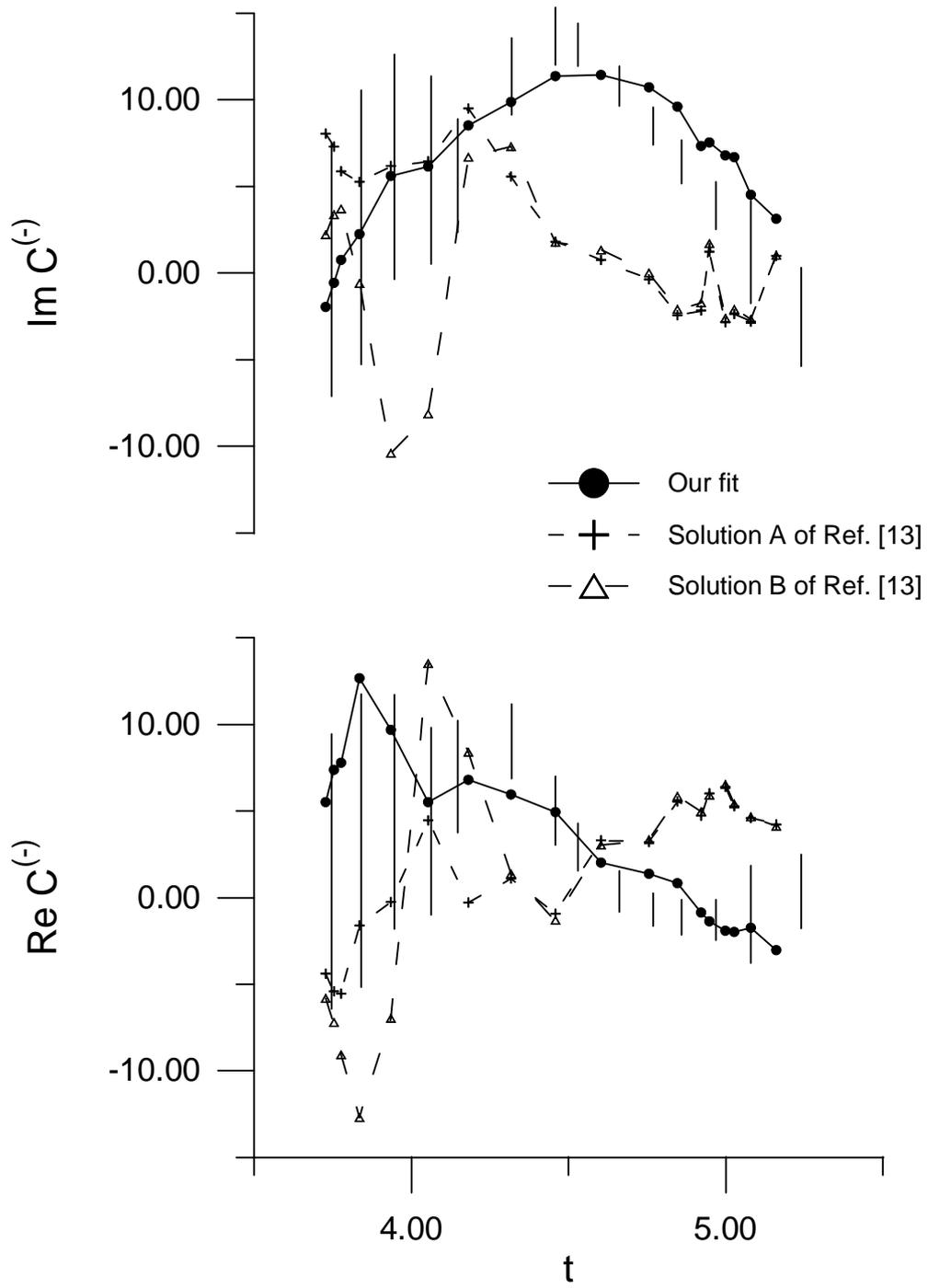
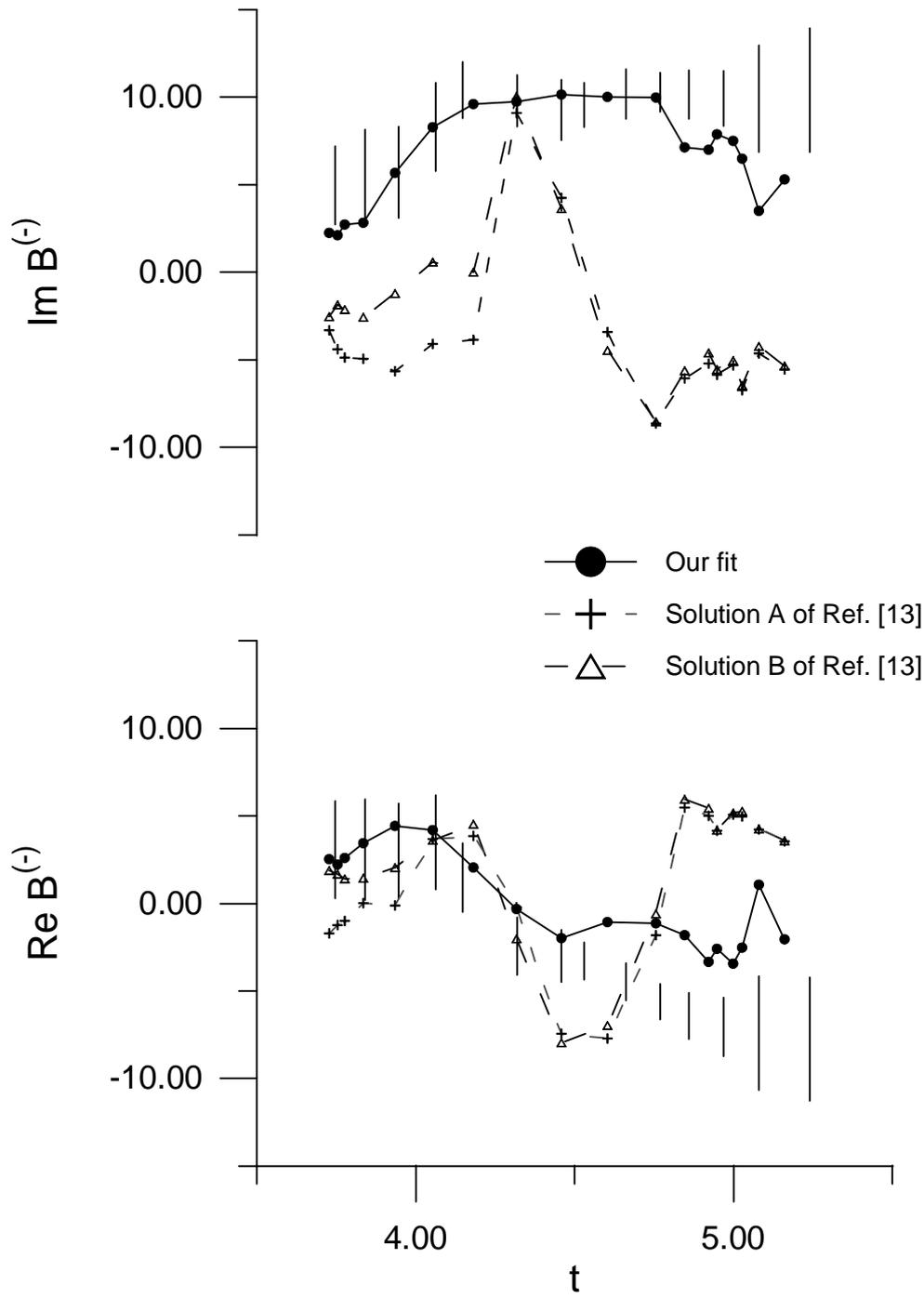
\begin{figure}[tbp]
\caption{Fits to the real and imaginary parts of $B^{\pm}$ and $C^{\pm}$ as
functions of $t$ at a fixed value of the hyperbolic parameter $\xi =
-0.04572 \;{\rm {GeV ^{2}}}$. The range of values for a given value of $t$
from ref. \protect\cite{ref8} are shown by the vertical lines, our fits are
shown by solid lines and the predictions of ref. \protect\cite{ref13} by
dashed lines. All values are in GeV units. (a) $C^{(+)}$, (b) $B^{(+)}$, (c) 
$C^{(-)}$, (d) $B^{(-)}$.}
\label{fig3}
\end{figure}

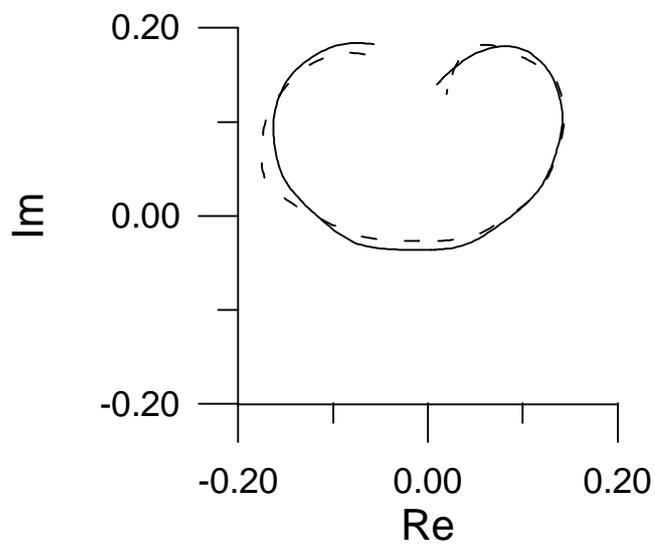
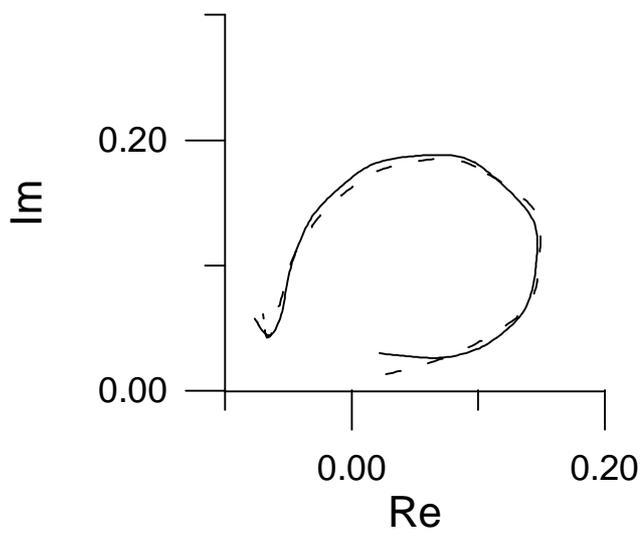
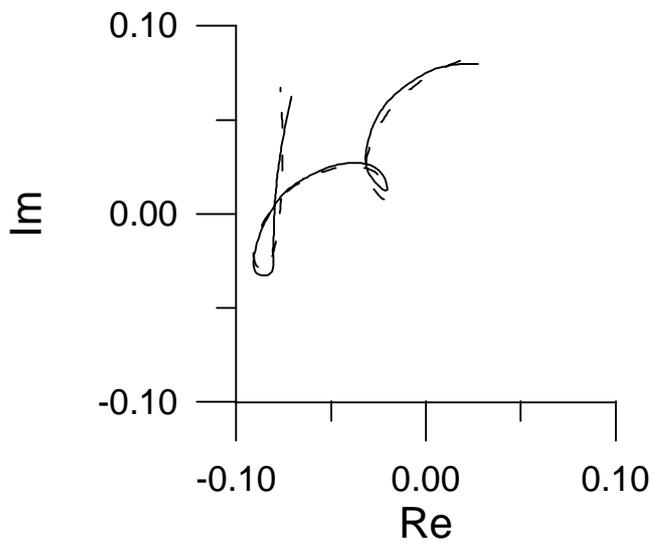
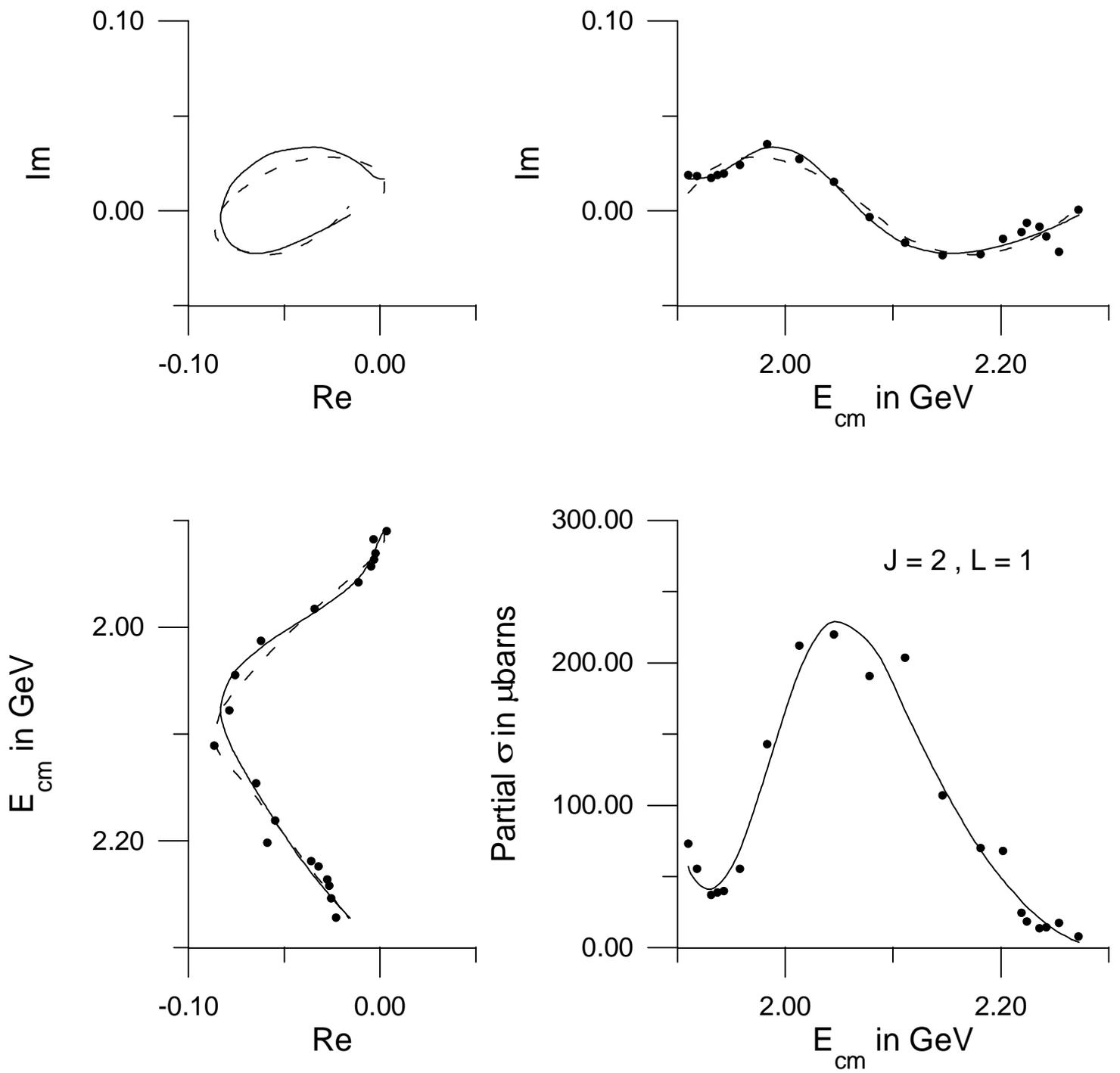
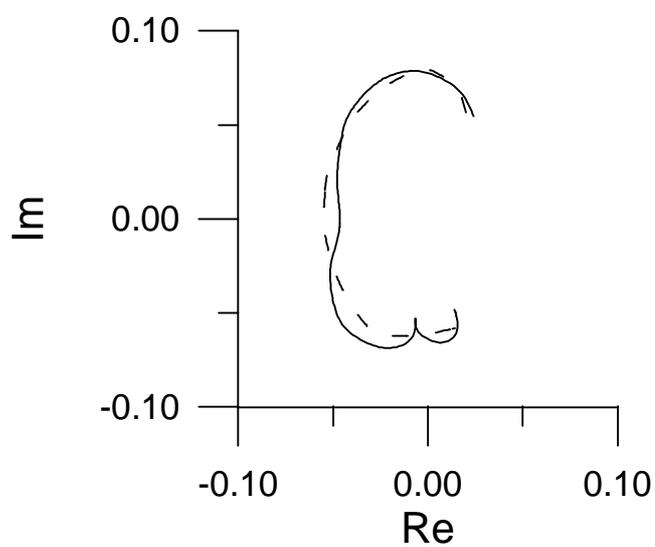
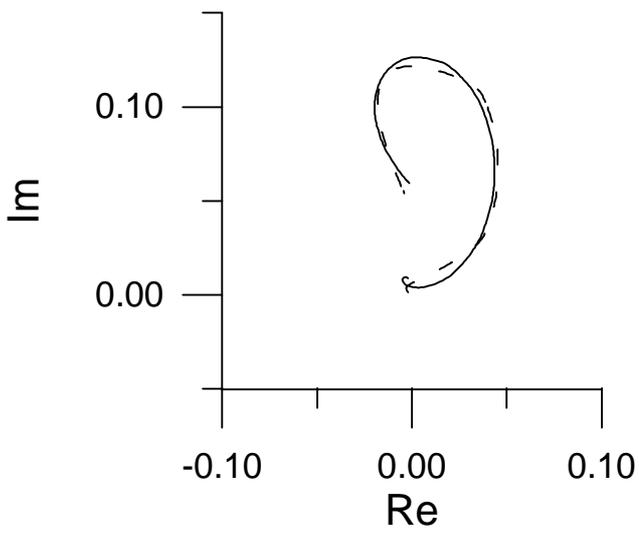
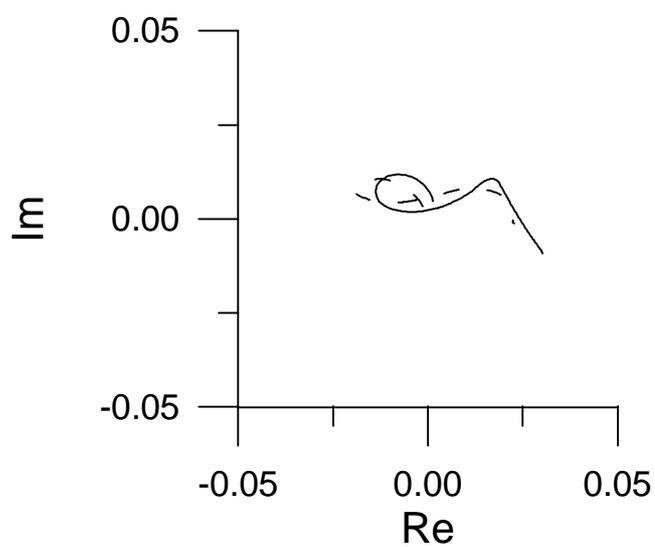
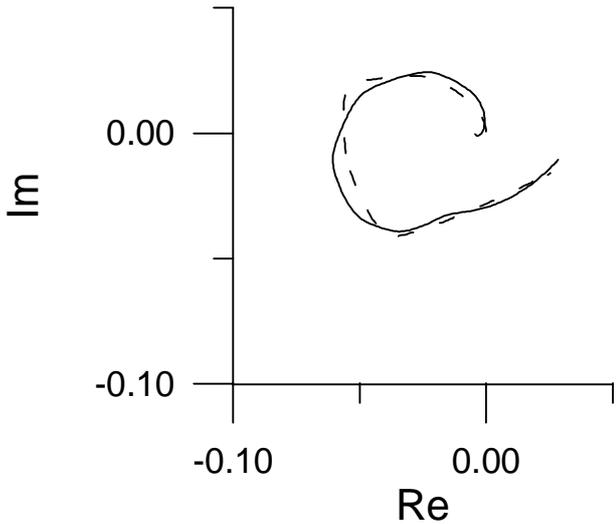
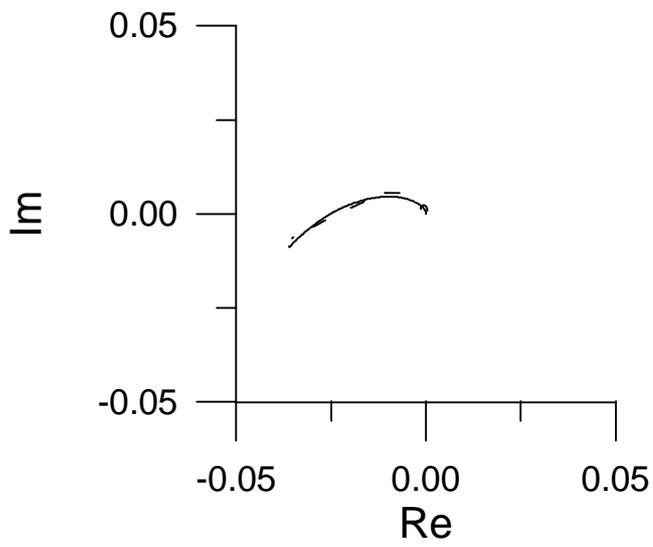
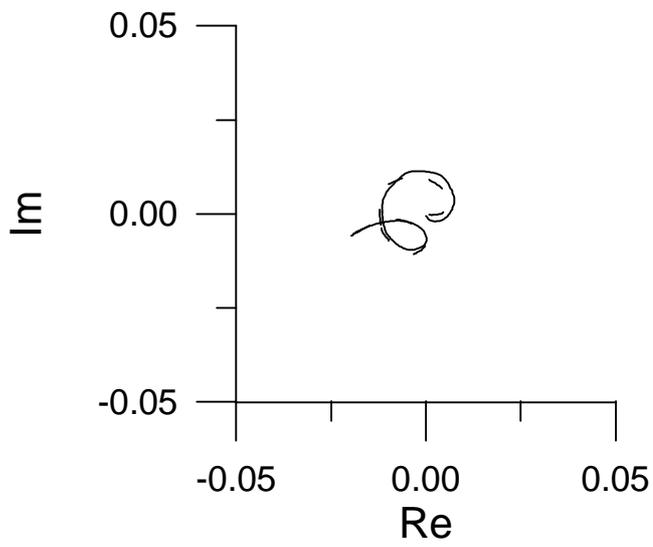
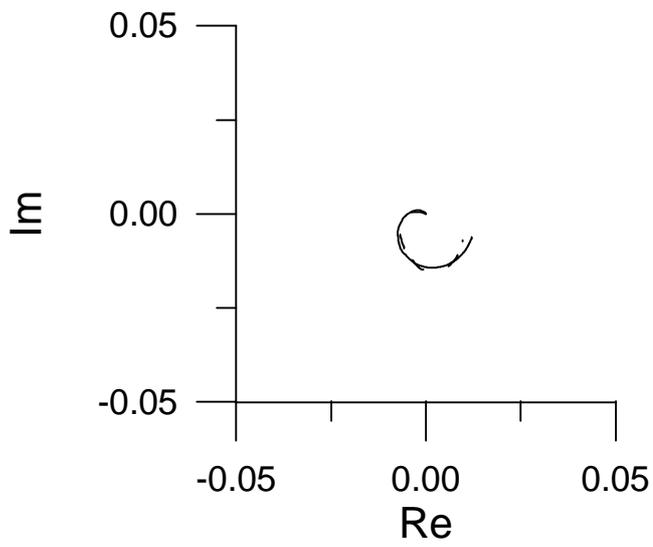
\begin{figure}[tbp]
\caption{Results of fitting the JL basis partial wave helicity amplitudes as
functions of energy with parametric forms incorporating partial-wave
analyticity based on eq.\ref{eq10a} shown by the solid lines, and with
simple resonance plus background forms, eq.\ref{eq12}, shown by the dashed
lines. The solid circles show the results of the single energy analyses. (a) 
$J=0, L=1$, (b) $J=1, L=0$, (c) $J=1, L=2$, (d) $J=2, L=1$, (e) $J=2, L=3$,
(f) $J=3, L=2$, (g) $J=3, L=4$, (h) $J=4, L=3$, (i) $J=4, L=5$, (j) $J=5,
L=4 $, (k) $J=5, L=6$.}
\label{fig4}
\end{figure}

One place where it may be possible to decide between the various solutions
is in the $\overline{p}p\rightarrow \pi ^{0}\pi ^{0}$ channel where dcs data
exist for only a limited momentum range. We show in Fig. \ref{fig2} our
predictions together with those of ref. \cite{ref13} for the differential
cross section and asymmetry in this channel at a lower momentum where there
are no measurements. In the same figure we also show similar predictions at
two higher momenta where measurements of the differential cross section
exist \cite{ref3}.

In conclusion, we summarise our overall findings. We have analysed data on $%
\overline{p}p\rightarrow \pi \pi $, including very accurate data in the $\pi
^{-}\pi ^{+}$ channel, using constraints from crossing symmetry and
analyticity to impose consistency of the solution with information on data
in the $\pi N\rightarrow \pi N$ channel. Overall, we have found evidence for
a rich spectrum of resonances whose couplings, with one or two exceptions,
are in the range expected for normal daughter meson states with a $\overline{%
q}q$ structure. The exceptions are in low J states and have couplings far
larger than expected. This may indicate the presence of more complex
multi-quark structures with large couplings to the $\overline{N}N$ channel
as expected in some models \cite{ref15}.

\newpage 

\newpage 
\begin{table}[tbp]
\caption{Values of $\chi ^{2}$ per data point obtained in fits to
differential cross-section and asymmetry data for $\bar{p} p \rightarrow
\pi^{-} \pi^{+}$, differential cross-section data for $\bar{p} p \rightarrow
\pi^{0} \pi^{0}$ and amplitude data}
\label{table1}%
\begin{tabular}{dddd}
Momentum (MeV/c) & $\pi ^{-}\pi ^{+}$ & $\pi ^{0}\pi ^{0}$ & Amplitudes \\ 
\hline
360 & 1.89 &  & 0.03 \\ 
404 & 1.29 &  & 0.03 \\ 
467 & 1.76 &  & 0.05 \\ 
497 & 2.32 &  & 0.04 \\ 
523 & 1.68 &  & 0.05 \\ 
585 & 1.96 &  & 0.05 \\ 
679 & 1.70 &  & 0.04 \\ 
783 & 1.49 &  & 0.03 \\ 
886 & 1.05 &  & 0.03 \\ 
988 & 1.40 &  & 0.07 \\ 
1089 & 1.40 &  & 0.10 \\ 
1190 & 1.64 & 0.28 & 0.09 \\ 
1291 & 1.76 & 0.66 & 0.15 \\ 
1351 & 2.16 & 0.12 & 0.22 \\ 
1400 & 1.91 & 0.18 & 0.18 \\ 
1416 & 1.69 & 0.36 & 0.17 \\ 
1449 & 1.80 & 0.23 & 0.26 \\ 
1467 & 1.75 & 0.20 & 0.31 \\ 
1500 & 1.46 & 0.41 & 0.38 \\ 
1550 & 1.66 & 0.35 & 0.35
\end{tabular}
\end{table}
\begin{table}[tbp]
\caption{Resonance masses and widths in units of GeV, obtained from fitting
partial-wave amplitudes as functions of energy, together with the values of
the product of branching ratios $B_J \equiv B(R \rightarrow \pi \pi)B(R
\rightarrow \bar{N}N).$}
\label{table2}%
\begin{tabular}{dddd}
$J$ & Mass & Width & $B_J$ \\ 
\hline
0 & 1.94, 1.95 & 0.16 & 0.18, 0.19 \\ 
1 & 1.96, 1.97 & 0.13, 0.14 & 0.053, 0.056 \\ 
2 & 1.93 & 0.15 & 0.013 \\ 
3 & 2.02 & 0.26, 0.22 & 0.013, 0.028 \\ 
4 & 2.02, 2.00 & 0.26, 0.14 & 0.013, 0.004 \\ 
5 & 2.19 & 0.22 & 0.004, 0.001 
\end{tabular}
\end{table}

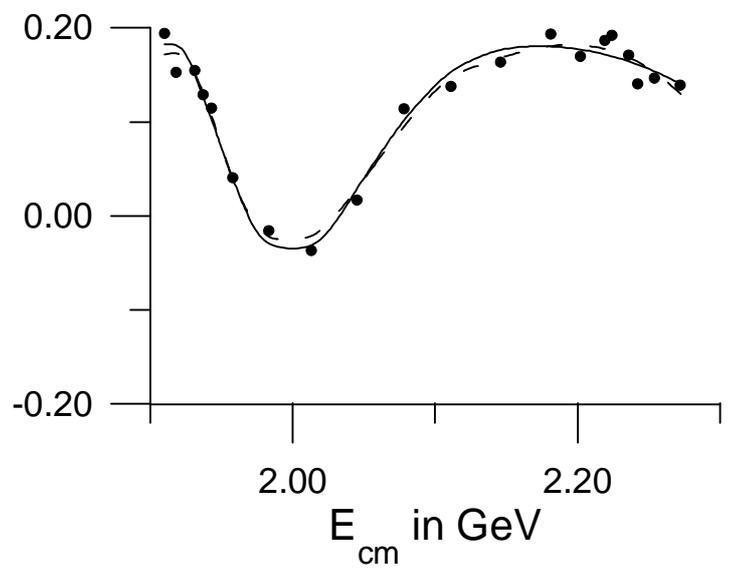
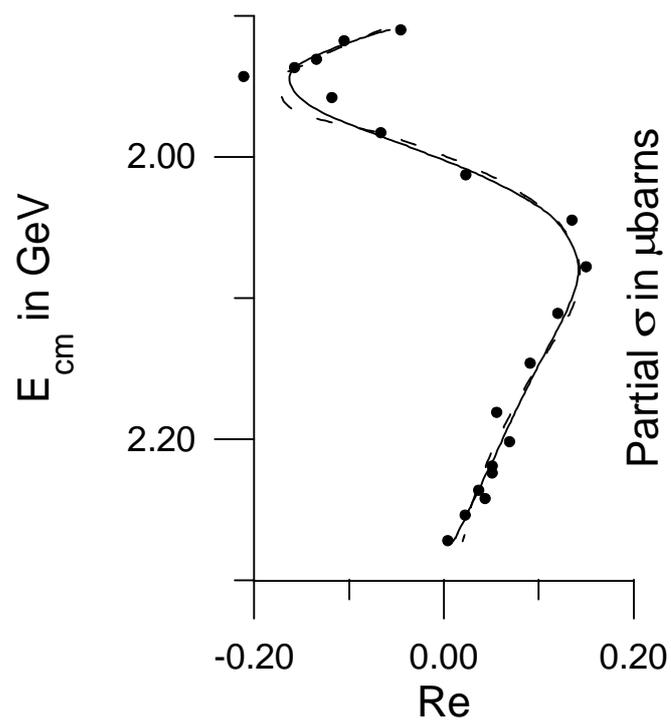
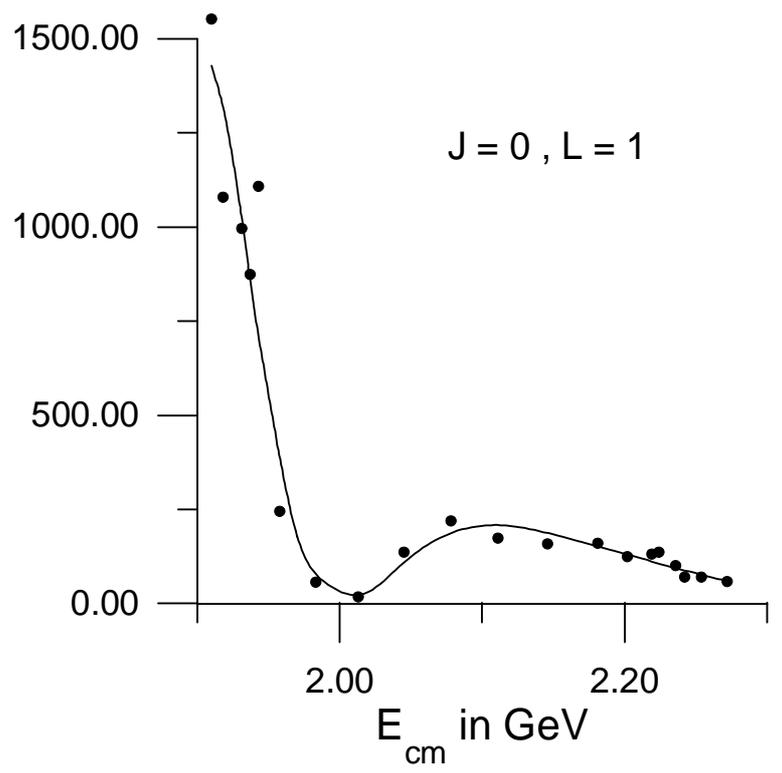
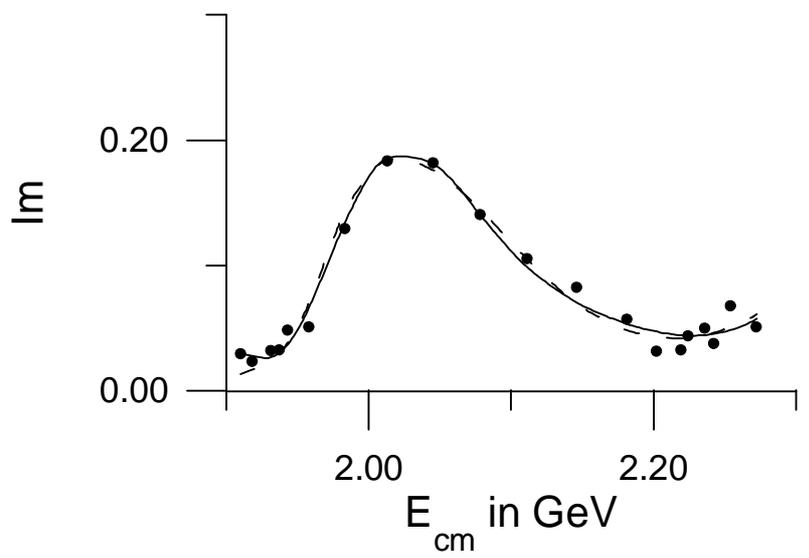
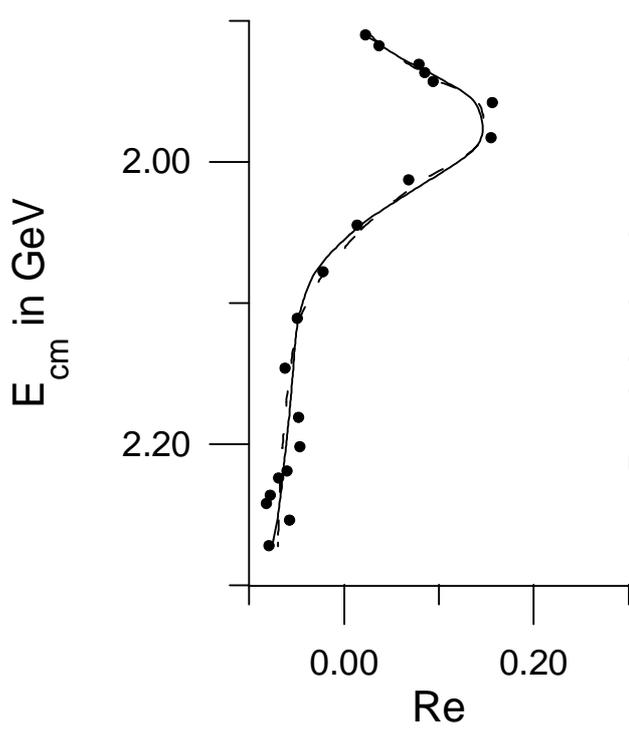
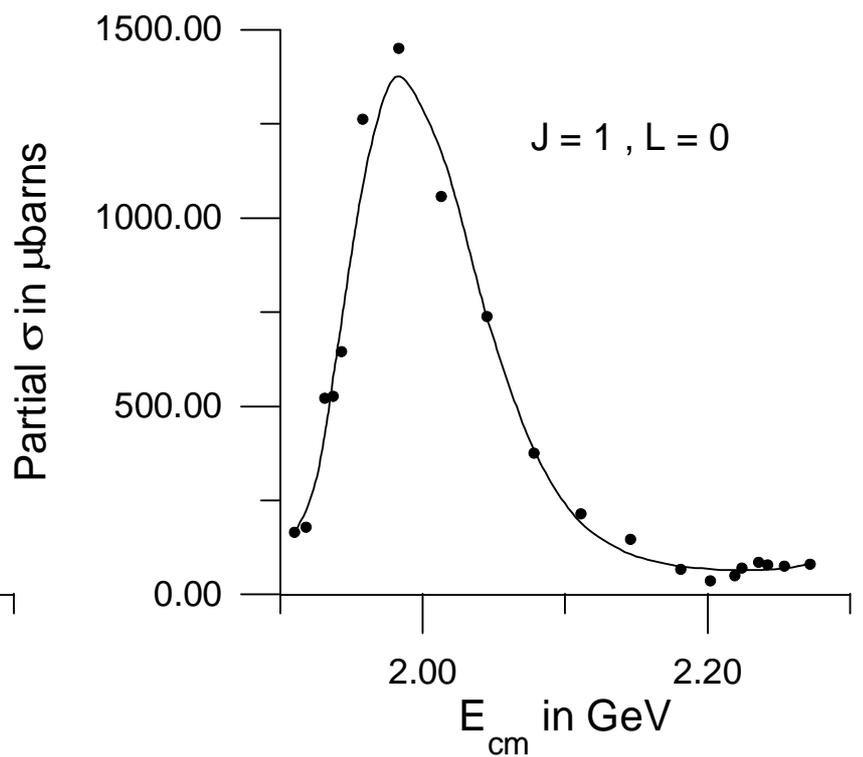
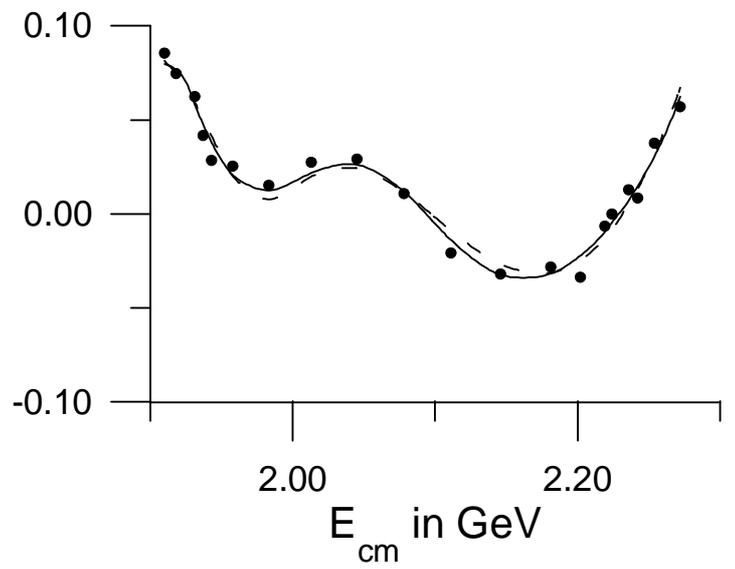
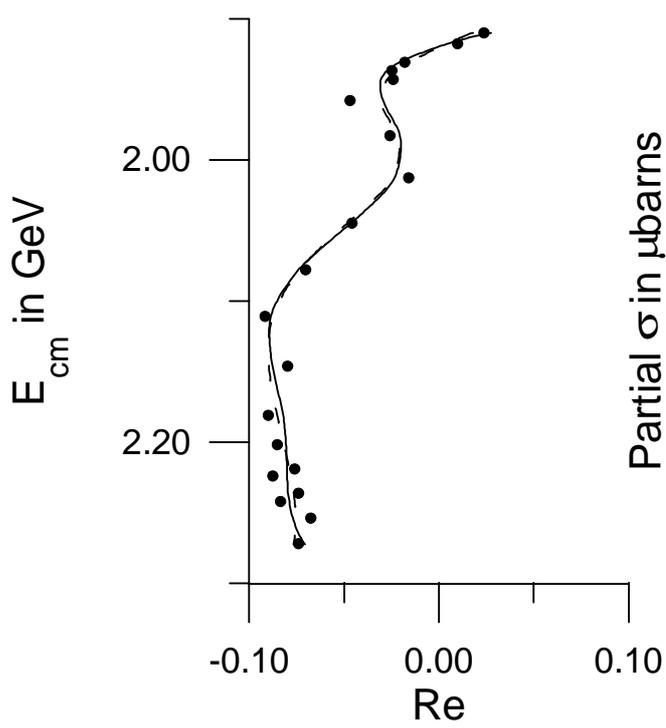
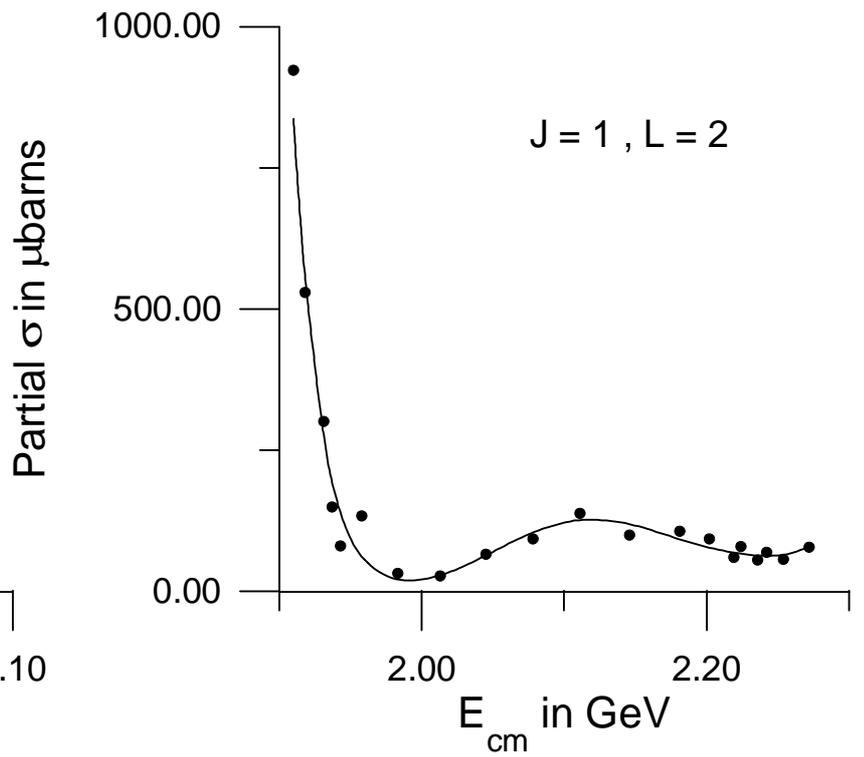
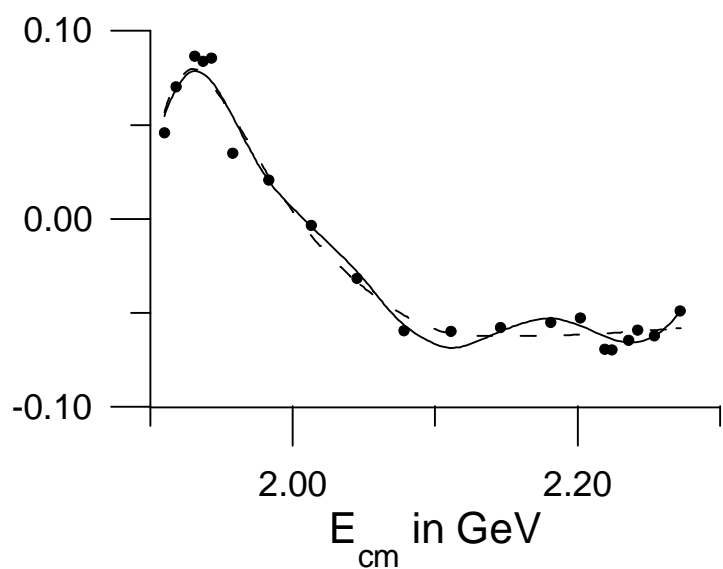
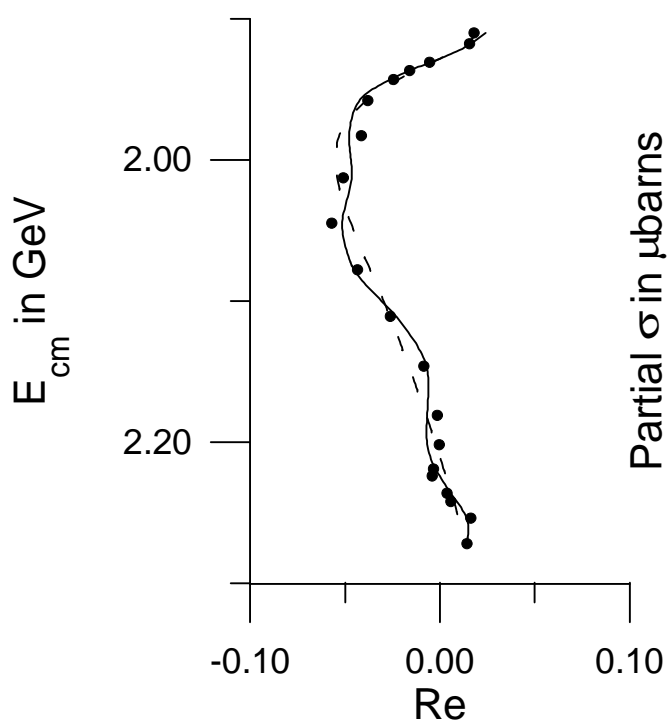
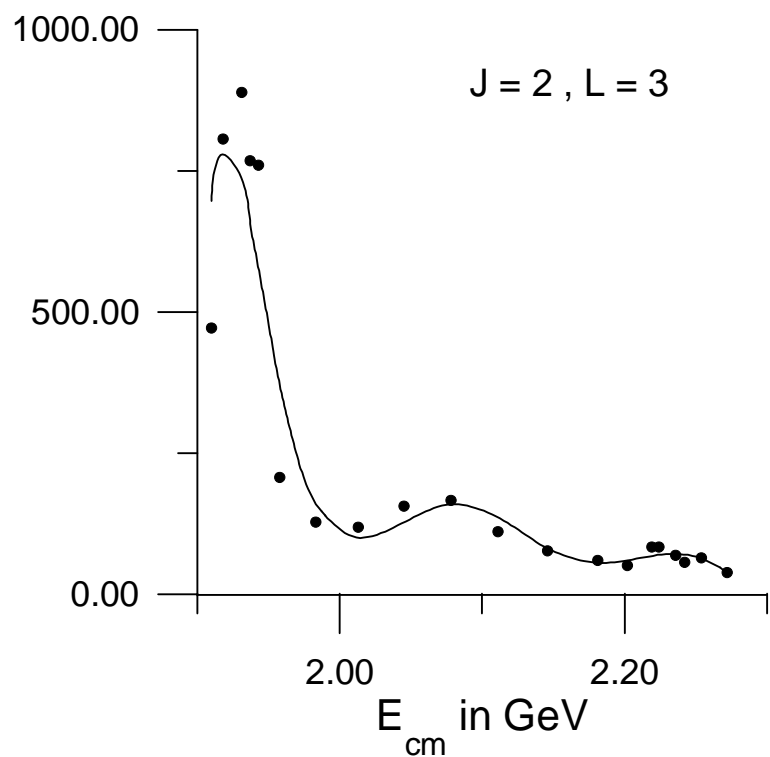
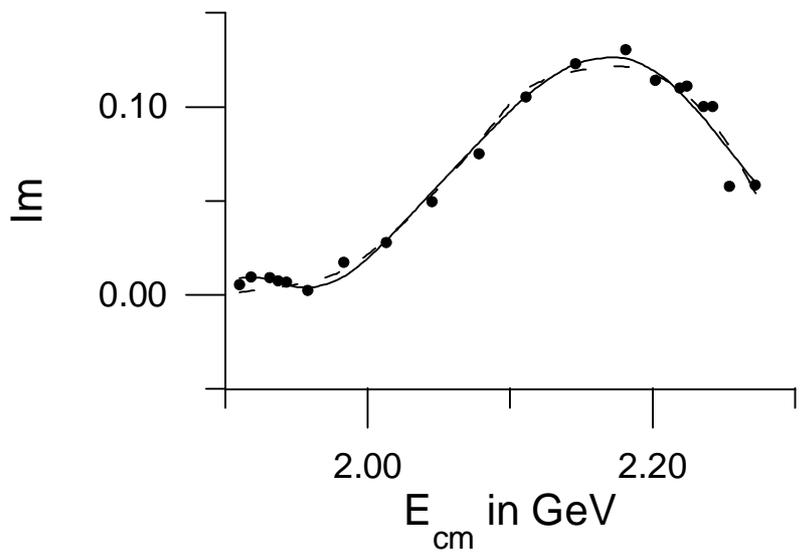
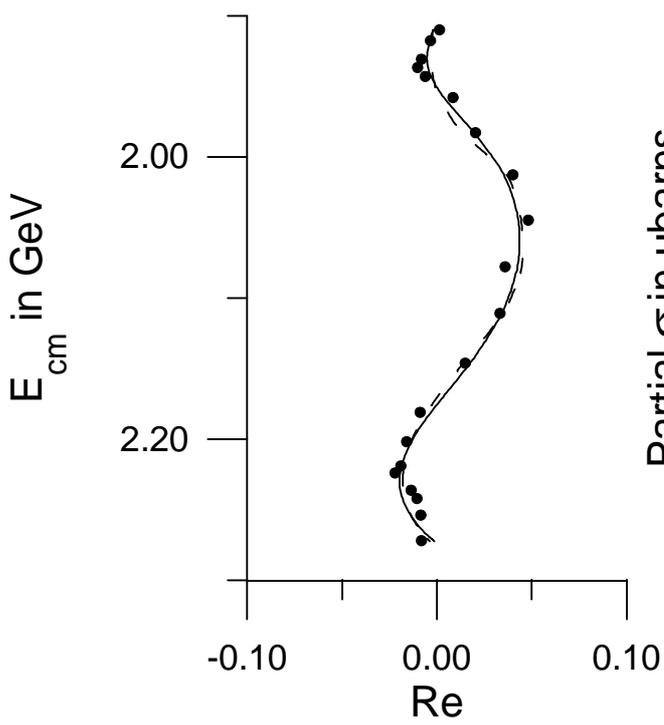
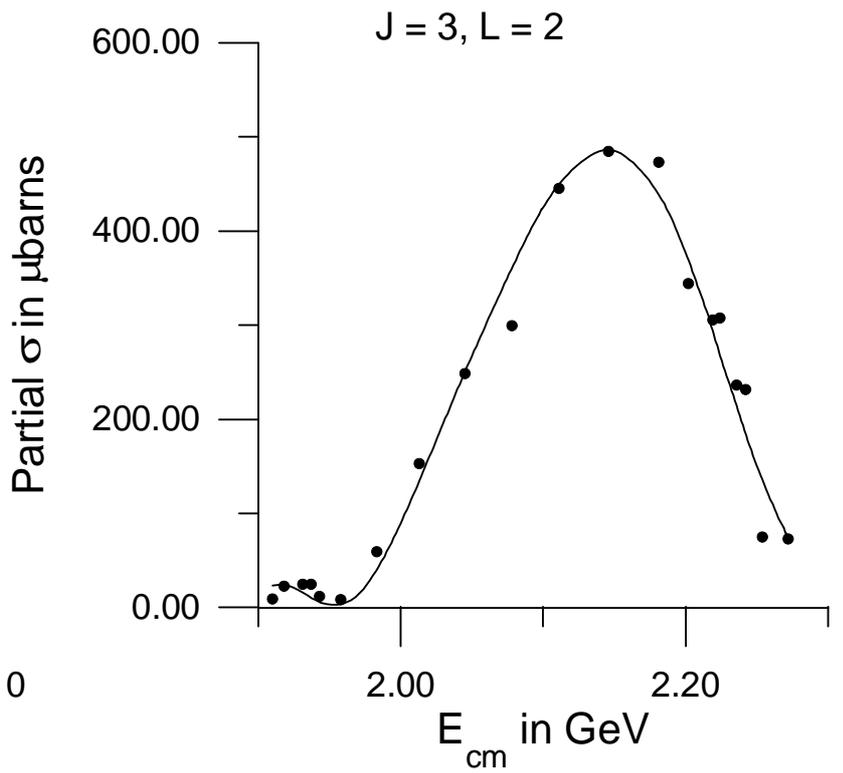
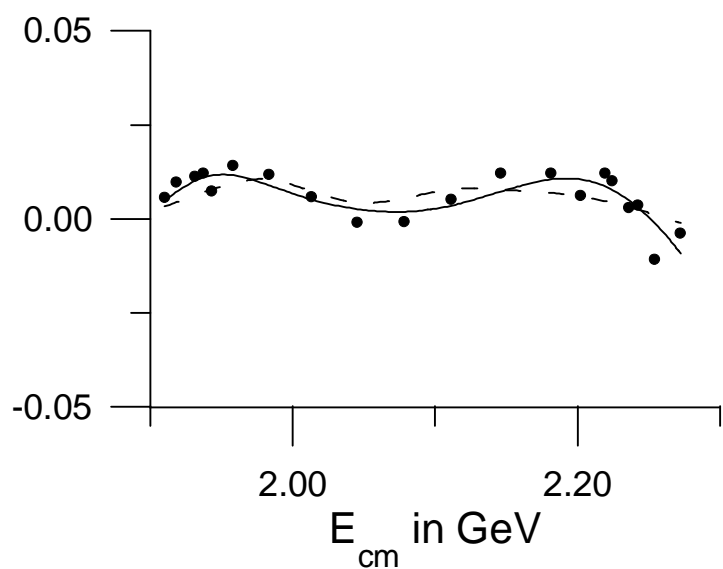
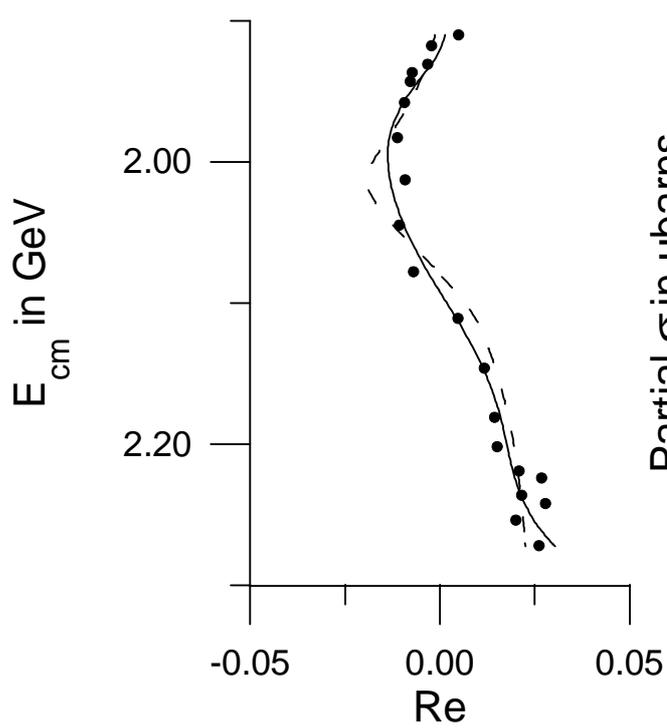
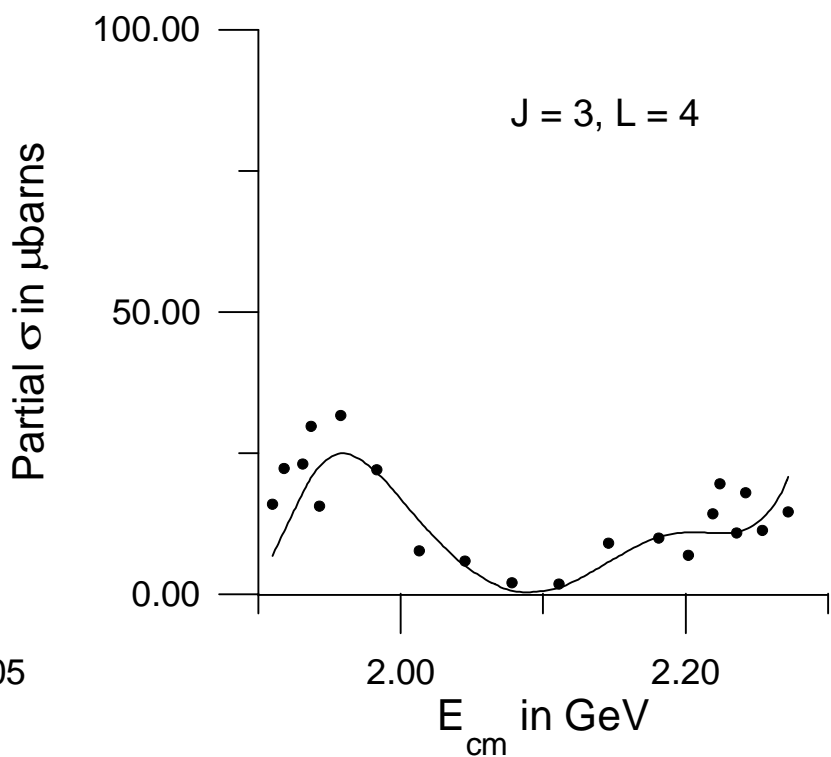
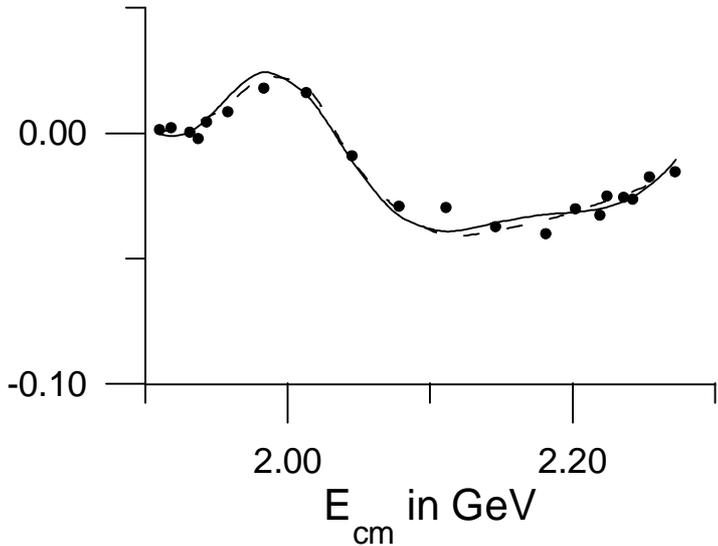
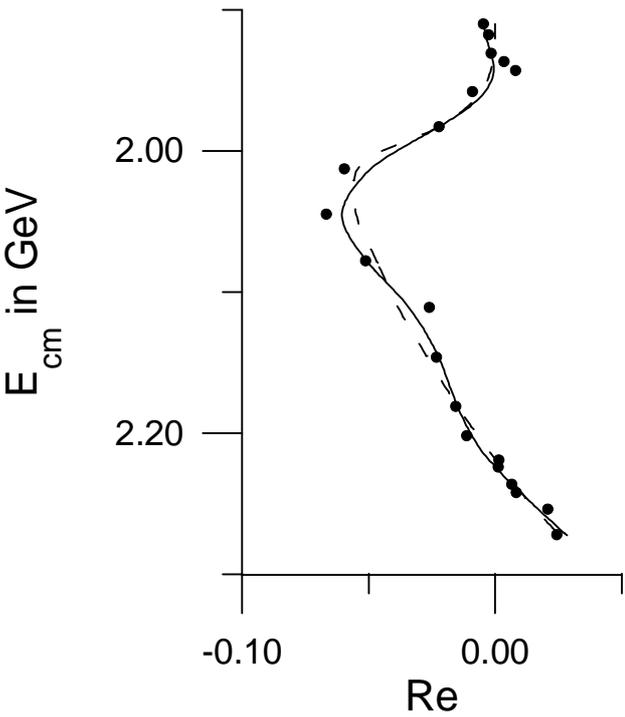
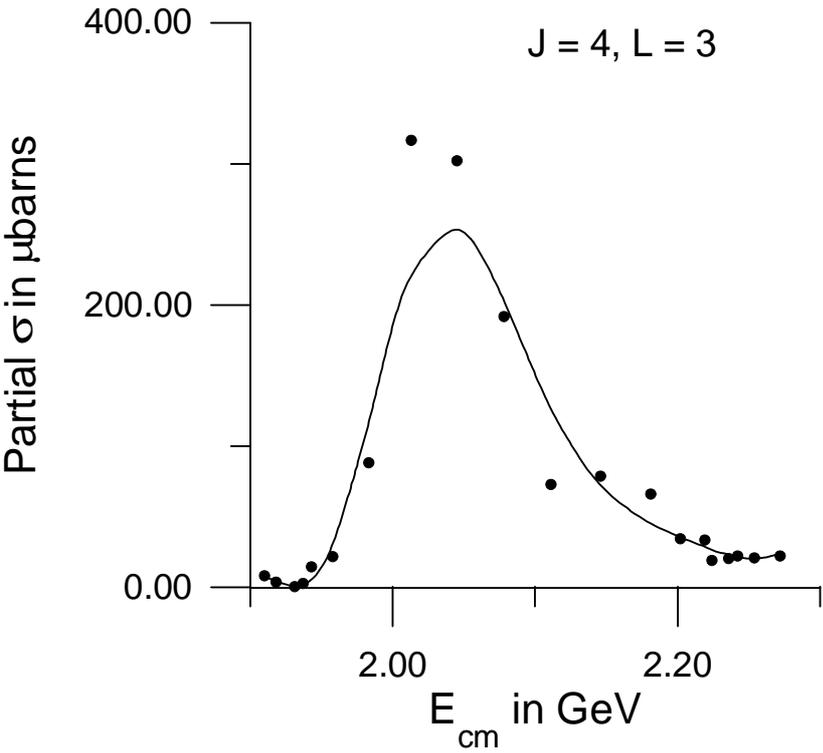
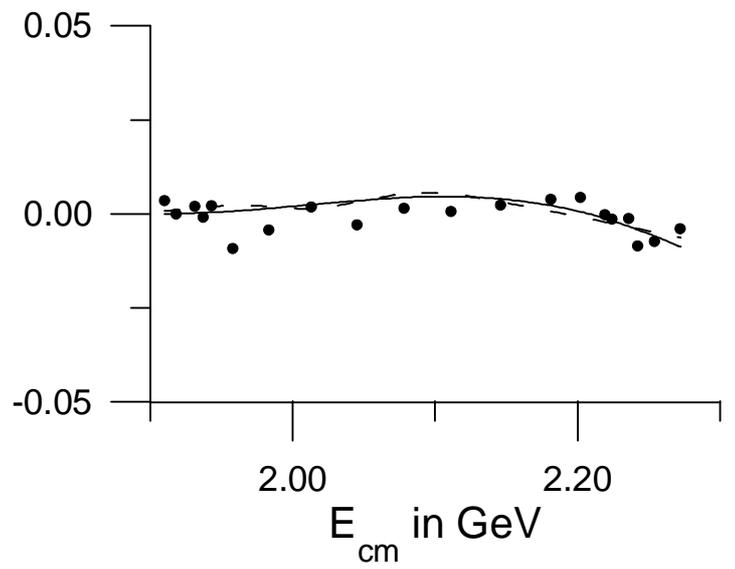
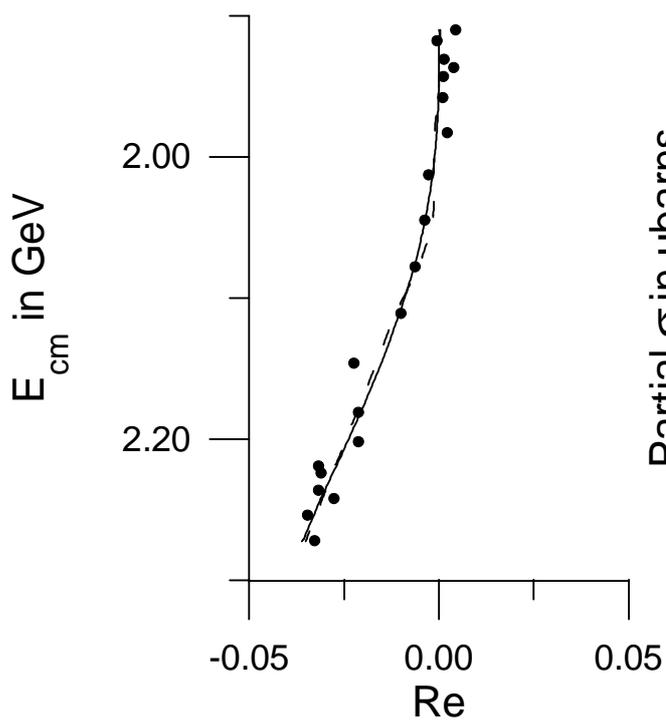
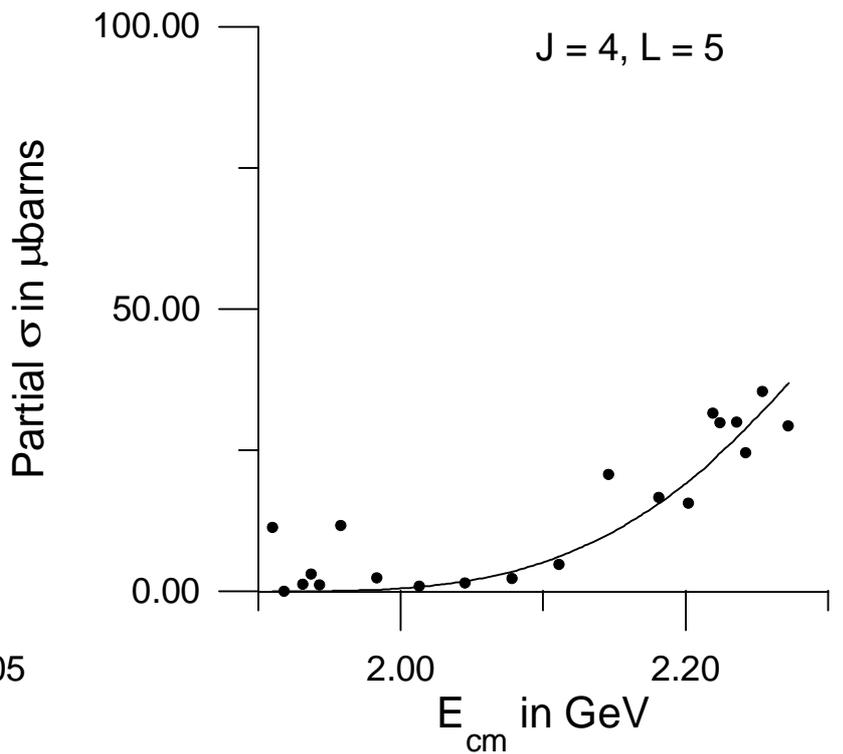
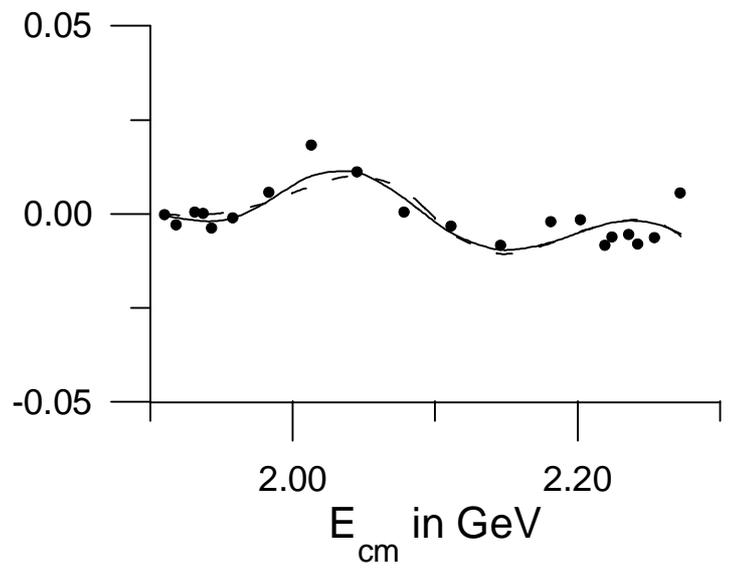
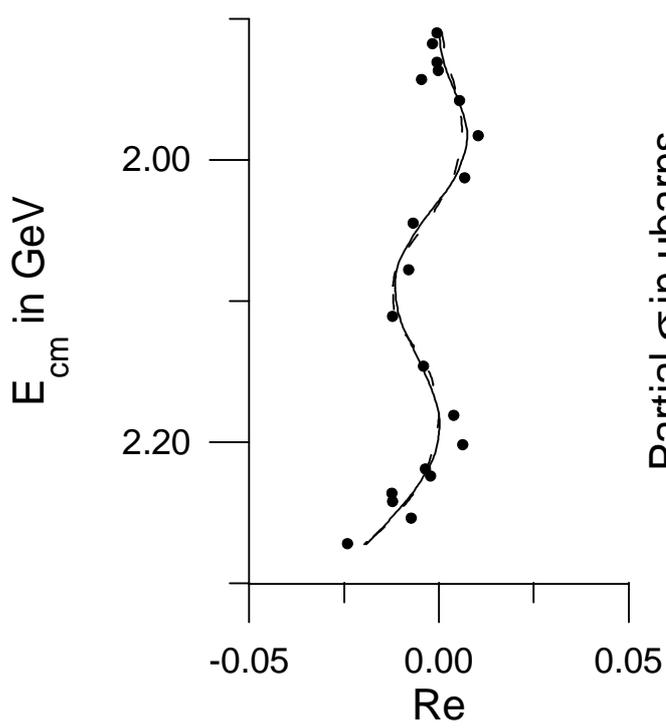
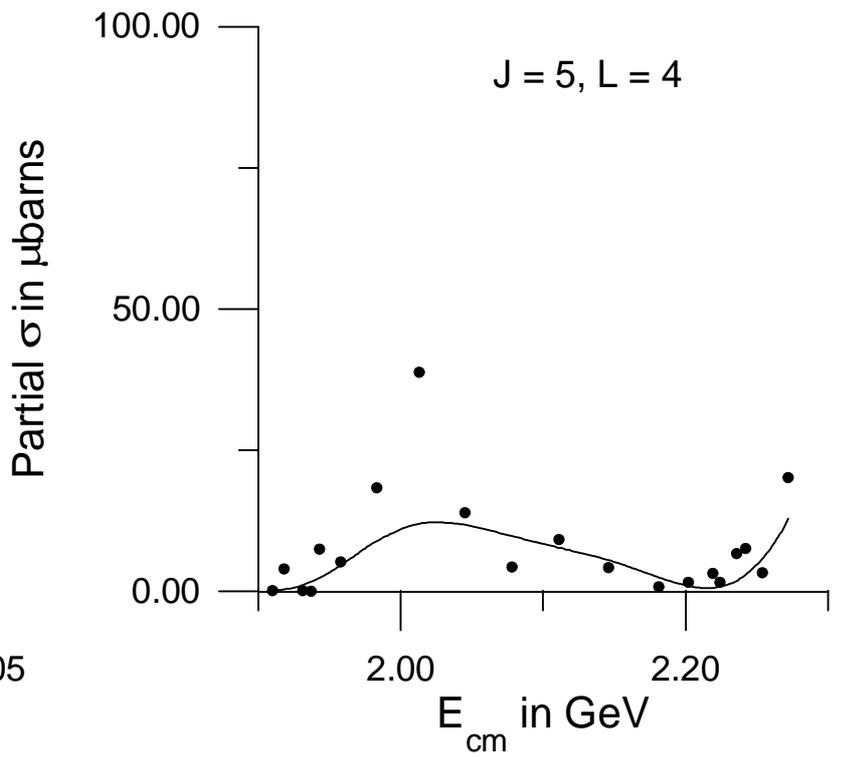
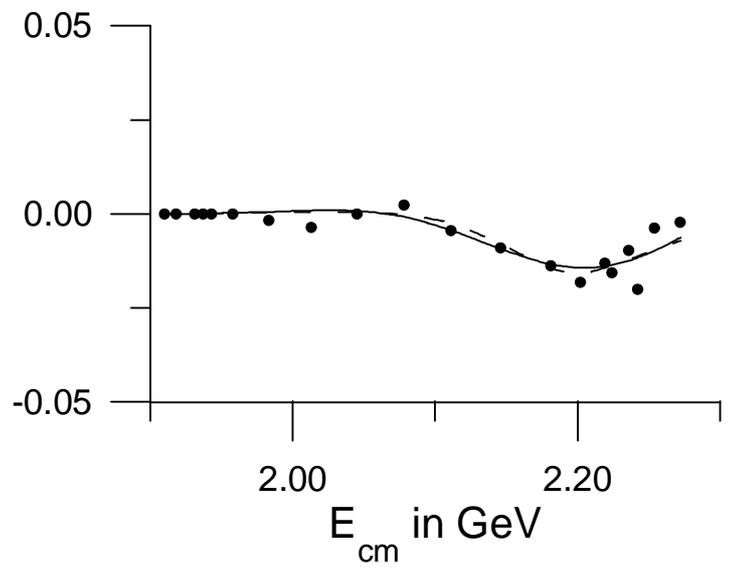
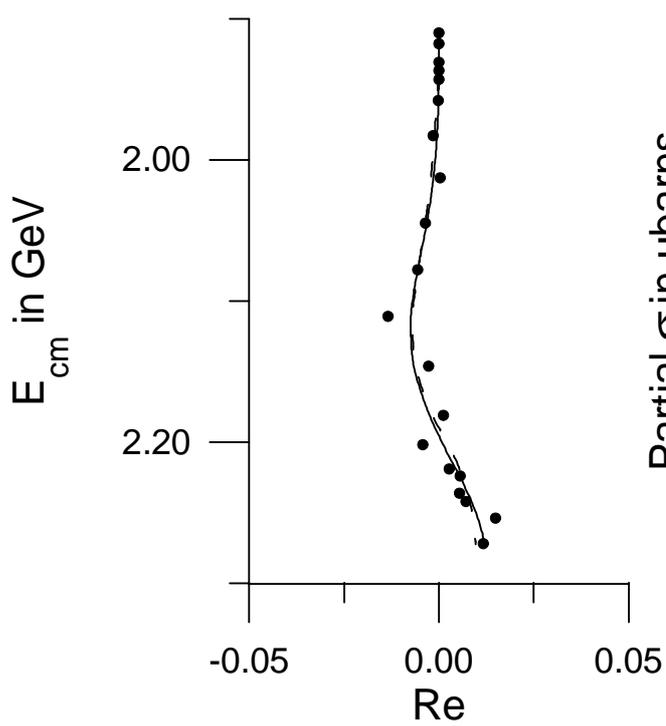
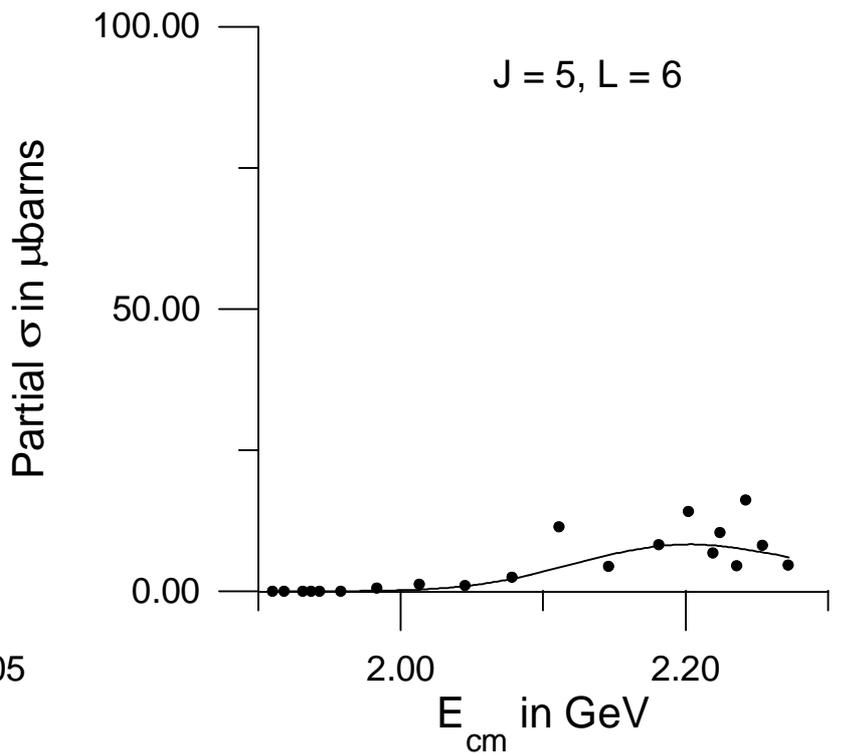
\end{document}